\documentclass[reprint, amsmath, amssymb, aps, prd, nofootinbib]{revtex4-2}

\usepackage{graphicx}
\usepackage{bm}
\usepackage{hyperref}
\usepackage{physics}
\usepackage[caption=false]{subfig}
\usepackage{orcidlink}

\newcommand\params{\ensuremath{\vec{\theta}}}

\newcommand{\bham}{Institute for Gravitational Wave Astronomy \& School of Physics and Astronomy, University of Birmingham, Edgbaston, Birmingham B15 2TT, UK}

\begin{document}

\title{Searching for a Ringdown Overtone in GW150914}

\author{Eliot Finch 
\orcidlink{0000-0002-1993-4263}}
\email{efinch@star.sr.bham.ac.uk}
\affiliation{\bham}

\author{Christopher J. Moore
\orcidlink{0000-0002-2527-0213}}
\email{cmoore@star.sr.bham.ac.uk}
\affiliation{\bham}

\date{\today}

\begin{abstract}
    We reanalyze the GW150914 post-merger data searching for quasinormal modes beyond the fundamental, quadrupolar mode.
    There is currently an ongoing disagreement in the literature about whether, and to what extent, the data contains evidence for a quasinormal mode overtone. 
    We use a frequency-domain approach to ringdown data analysis that was recently proposed by the authors. 
    Our analysis has several advantages compared to other analyses performed mainly in the time domain; in particular, the source sky position and the ringdown start time are marginalized over (as opposed to simply being fixed) as part of a Bayesian ringdown analysis.
    We find tentative evidence for an overtone in GW150914, but at a lower significance than reported elsewhere.
    Our preferred analysis, marginalizing over the uncertainty in the time of peak strain amplitude, gives a posterior on the overtone amplitude peaked away from zero at $\sim 1.8\sigma$.
\end{abstract}

\maketitle

\section{Introduction}\label{sec:introduction}

The gravitational-wave (GW) signals from binary coalescences provide a unique opportunity to study gravity in the strong-field and dynamical regimes.
Of particular interest here is the signal from the final stages of a binary black-hole (BH) merger, known as the ringdown, which is associated with the remnant BH settling into its final state.
The ringdown contains a superposition of exponentially damped oscillations, known as quasinormal modes (QNMs), with a discrete set of (complex) frequencies.
Identifying these frequencies allows us to measure properties of the remnant and also provides a particularly clean way to test general relativity (GR) and the Kerr metric; this procedure is known as black hole spectroscopy \cite{Dreyer:2003bv}.
QNMs have now been identified in a few tens of binary BH merger signals in the most recent GW catalogs~\cite{LIGOScientific:2020tif, LIGOScientific:2021sio}.

However, the very first GW event, GW150914 \cite{LIGOScientific:2016aoc}, remains probably the best candidate for studying the ringdown.
This is a result of several factors, including its large signal-to-noise ratio (SNR) of $\rho\sim 24$ and its total mass of $M\sim 70\,M_\odot$ which places the merger and ringdown in the center of the LIGO \cite{LIGOScientific:2014pky} sensitive frequency band at $\sim 200\,\mathrm{Hz}$. 
Additionally, GW150914 is by now the most well-studied GW event and therefore the signal and the properties of the noise in the surrounding data are extremely well understood.

The first tests of GR performed using GW150914 included an investigation of the ringdown~\cite{LIGOScientific:2016lio}. 
The ringdown signal, after a fixed starting time $t_0$, was modeled using a single damped sinusoid; the parameters of which were checked for consistency with the predicted least-damped QNM of the remnant BH.
This first attempt at a ringdown analysis was performed using the standard Whittle frequency-domain log-likelihood~\cite{Whittle:1957}, commonly used in GW data analysis.
The ringdown was isolated by choosing a lower limit of $\sim 130\, \mathrm{Hz}$ in the frequency integral, effectively cutting the data mid-signal.
This approach suffers from several shortcomings. 
Firstly the frequency-domain cut at $\sim 130\, \mathrm{Hz}$ only approximately separates the ringdown from the early-time signal due to the breakdown of the stationary phase approximation near merger. 
Secondly the nonzero amplitude at the start of the signal model breaks the assumption of circularity for the Fourier transform, thereby introducing contamination in the form of spectral leakage. 
Therefore, this approach does not scale well to higher SNRs where noise will no longer dominate over the systematic errors introduced by the sharp frequency-domain cut.
Despite these drawbacks, this approach was successfully used in Ref.~\cite{LIGOScientific:2016lio} to identify the fundamental QNM in the GW150914 signal.

Since this initial attempt, several groups have developed new time-domain frameworks specifically for ringdown analyses \cite{Carullo:2019flw, Isi:2019aib, Capano:2021etf}.
The principle motivation for working in the time domain is that it is easy to impose sharp cuts on the data at specific times (without any spectral leakage) and to analyze only data after a chosen start time (see Ref.~\cite{Isi:2021iql} for details of time-domain analysis methods).
These approaches have also enabled going beyond the fundamental mode. 
Generically, the ringdown can be modeled as a superposition of QNMs with complex frequencies $\omega_{\ell m n} = 2\pi f_{\ell m n} - i/\tau_{\ell m n}$, labeled with angular indices $\ell\geq 2$, $\abs{m}\leq\ell$, and an overtone index $n \geq 0$ [the fundamental mode has $(\ell, \abs{m}, n) = (2, 2, 0)$].
Detecting additional QNMs beyond the fundamental increases the scientific potential of ringdown studies, especially for fundamental tests of the Kerr metric, the no-hair theorems, and the BH area law \cite{Dreyer:2003bv, Berti:2005ys, Gossan:2011ha, Brito:2018rfr, Carullo:2019flw, Isi:2019aib, Isi:2020tac}.

An early application of the time-domain framework was in Ref.~\cite{Isi:2019aib}, where Isi et al. claimed a detection of the first overtone of the fundamental QNM in the GW150914 signal [that is, the $(2, 2, 1)$ mode]. 
This was quickly followed by a separate detection claim of the $(3,3,0)$ harmonic mode in the signal of the $\sim 150M_\odot$ binary merger GW190521~\cite{LIGOScientific:2020iuh} by Capano et al.~\cite{Capano:2021etf} (this was done using an equivalent formulation of the time-domain method, although expressed in the frequency domain).
The claimed detection of an overtone was made possible partly because, compared to earlier studies, the authors chose to use an earlier start time for the ringdown; this was motivated by contemporary numerical relativity studies~\cite{Giesler:2019uxc} (see also Refs.~\cite{Bhagwat:2019dtm, Ota:2019bzl, Cook:2020otn, JimenezForteza:2020cve, Dhani:2020nik, Finch:2021iip, Forteza:2021wfq, Dhani:2021vac, MaganaZertuche:2021syq}) that demonstrated that when overtones are included the ringdown can be considered to start as early as the time of peak strain amplitude. 

However, a recent paper by Cotesta et al.~\cite{Cotesta:2022pci} reanalyzed the GW150914 signal using very similar methods and found no significant evidence for an overtone.
It was also suggested that the earlier detection claims of Ref.~\cite{Isi:2019aib} were noise dominated.
(This prompted a response from Isi et al.~\cite{Isi:2022mhy} where they restated their claim to have detected an overtone in GW150914.) 
Ref.~\cite{Bustillo:2020buq} also found weaker evidence for an overtone using an analysis method closer to that of Ref.~\cite{LIGOScientific:2016lio}.
Similarly, the claim in Ref.~\cite{Capano:2021etf} that a harmonic had been detected in GW190521 has also been debated and no evidence for a harmonic was found by Ref.~\cite{LIGOScientific:2021sio}.
Amid this confusion, it is particularly concerning that the supposedly identical analyses in Refs.~\cite{Isi:2019aib, Isi:2022mhy}, and \cite{Cotesta:2022pci} come to such different conclusions concerning which QNMs are in the data. 
Discrepancies of this sort risk jeopardizing the science that can be done using future ringdown observations.

These discrepancies highlight some of the difficulties inherent in time-domain ringdown analysis, where important choices (that affect the results) for fixed quantities such as the ringdown start time have to be made and care must be taken with the noise covariance estimation.
If ringdown studies are to be used to make precision measurements of BH properties or as reliable tests of GR we must first be able to make reliable and reproducible determinations of the QNM content.
This is also not a problem that will be removed in the future with observations at higher SNR. Even if an event has a higher SNR that is sufficient for a clear detection of the first QNM overtone, the focus will then simply shift to trying to identify the next overtone (or else the next QNM harmonic) in the countably infinite ringdown sum \cite{Bustillo:2020buq}.

To complement the time-domain analysis frameworks, the authors recently proposed a new method for ringdown analyses which works in the frequency domain \cite{Finch:2021qph}.
A flexible sum of sine-Gaussian wavelets, truncated at the ringdown start time, is used to effectively marginalize over the inspiral-merger (i.e.\ pre-ringdown) part of the signal.
The model is completed by attaching this to the usual sum of QNMs which model the ringdown.
No continuity is enforced between the two parts of the model in order to keep the ringdown inference independent from the rest of the signal.
However, we find the continuity is effectively learned from the data, and any remaining discontinuities disappear entirely when the signal is ``whitened’’ according to the instrumental noise.
In a particular limit, this approach can be shown to be formally equivalent to the time-domain analyses described above.
However, this frequency-domain approach can be generalized and offers several advantages over time-domain approaches:
well-established GW data analysis methods and pipelines can be used (which are all built in the frequency domain), 
the inspiral-merger data informs the noise estimation at the start of the ringdown (improving parameter estimation accuracy), 
and the ringdown start time and the source sky position can be easily treated as free parameters and marginalized over as part of a Bayesian analysis (instead of being fixed).
We note, however, that (as discussed in Ref.~\cite{Finch:2021qph}) a narrow and informative prior on the ringdown start time must be used.
Reweighting techniques can be employed to investigate different ringdown start time prior choices computationally efficiently in post processing (see Sec.~\ref{subsec:reweighting}) obviating the need for the large number of analyses performed in \cite{Cotesta:2022pci, Isi:2022mhy}.

In this paper the new frequency-domain method is applied to reanalyzing the ringdown of GW150914 paying particular attention to the presence (or absence) of an overtone. 
We perform analyses with and without an overtone and investigate different choices of the ringdown start time. 
We also perform additional analyses with varying data sampling frequencies and integration limits to verify the stability of our results. Finally, a mock injection study into real detector noise is also performed to further assess the significance of any overtone detection.
Sec.~\ref{sec:analysis} describes the signal model, the data, and the analysis methods used in this paper.
Sec.~\ref{sec:results} presents our main results including posteriors on the remnant BH properties and overtone amplitude, and Bayes' factors for the overtone model.
The results are discussed further in Sec.~\ref{sec:discussion}.
Throughout this paper we make use of natural units where $G=c=1$.

All data products and plotting scripts used to make the figures in this paper are made publicly available at Ref.~\cite{finch_eliot_zenodo}.

\section{Methods}\label{sec:analysis}

This section briefly describes the frequency-domain method for analyzing BH ringdowns introduced in Ref.~\cite{Finch:2021qph}:
the wavelet-ringdown model is described in Sec.~\ref{sec:model}; the data, likelihood and priors are described in Sec.~\ref{sec:details}; and our approach for dealing with changes to the ringdown start time is described in Sec.~\ref{subsec:reweighting}.

\subsection{Wavelet-Ringdown Model}\label{sec:model}

Our model consists of two parts: one for early times before $t_0$ which is referred to here as the \emph{inspiral-merger}, and another for the \emph{ringdown} after the start time $t_0$.

First, we describe the ringdown part of the model.
After a ringdown start time $t_0$, which is itself a parameter in the model, the model takes the form
\begin{align}
    h^\mathrm{R}(t) &= h_+^\mathrm{R}(t) + ih_\times^\mathrm{R}(t) \nonumber \\
    &= \sum_{n=0}^N A_n e^{-i[\omega_{22n}(t-t_0) + \phi_{n}]}, \quad t \geq t_0. \label{eq:ringdown_model}
\end{align}
Because our focus in this paper is on the presence of an overtone, we fix the angular indices to $\ell = m = 2$ and vary only the number of QNM overtones, $N$, in the model ($N$ is always taken to be either 0 or 1 in this paper). 
Note that the form of this equation differs slightly from Eq.~11 in Ref.~\cite{Finch:2021qph}. This is because the source inclination angle is fixed to be ``face-off'' (i.e.\ $\iota=\pi$).
In the notation of (for example) Refs.~\cite{Dhani:2020nik, Finch:2021iip, MaganaZertuche:2021syq}, this is equivalent to using the $\ell = -m = 2$ mirror modes. 
Or, in notation of Ref.~\cite{Isi:2021iql}, using an ellipticity of $\epsilon = -1$.
The complex QNM frequencies, $\omega_{\ell m n} = 2\pi f_{\ell m n} - i/\tau_{\ell m n}$, are functions of the remnant BH mass $M_f$ (detector frame) and dimensionless spin $\chi_f$.
Additionally, each QNM is further described by an amplitude, $A_{n}$, and a phase, $\phi_{n}$. 

Second, we describe the inspiral-merger part of the model.
This is modeled as a truncated sum of $W$ wavelets.
At early times the model takes the form
\begin{align} 
    h^\mathrm{IM}(t) &=  h_+^\mathrm{IM}(t) + ih_\times^\mathrm{IM}(t) \nonumber \\
    &= \sum_{w=1}^{W} \mathcal{A}_w \exp \Bigg[-2\pi i \nu_w(t-\eta_w) \label{eq:wavelets} \\
    &\hspace{2.46cm} - \qty(\frac{t-\eta_w}{\tau_w})^2 - i\varphi_w \Bigg], \quad t < t_0. \nonumber
\end{align}
Again, the minor differences in sign conventions compared to Ref.~\cite{Finch:2021qph} come from fixing the inclination angle to be face-off. 
The wavelets are each described by five parameters: $\mathcal{A}_w$ and $\varphi_w$ are the wavelet amplitudes and phases, $\tau_w$ are the wavelet widths, $\nu_w$ are the wavelet frequencies, and $\eta_w$ are the wavelet central times. 
In this paper we use $W=3$ (three wavelets) in our model.
This number was empirically found to be sufficient (see the appendix of Ref.~\cite{Finch:2021qph}, where the number of wavelets was varied for a GW150914-like injection).

The full signal model is given by discontinuously joining the inspiral-merger to the ringdown at $t_0$,
\begin{equation}
    h(t) = h^\mathrm{IM}(t) + h^\mathrm{R}(t).
\end{equation}

Finally, the detector response must be considered.
We project the waveform polarizations onto each interferometer (IFO) with the antenna patterns, $F^\mathrm{IFO}_{+,\times}$.
The detector response for each ${\mathrm{IFO}\in \{\mathrm{H}, \mathrm{L}\}}$ is given by
\begin{align} \label{eq:projection_antenna}
    h^\mathrm{IFO}(t) = F^\mathrm{IFO}_+(\alpha, \delta, \psi) ~ &h_+(t + \Delta t_\mathrm{IFO}) \nonumber \\
    + F^\mathrm{IFO}_\times(\alpha, \delta, \psi) ~ &h_\times(t + \Delta t_\mathrm{IFO}),
\end{align}
where $\alpha$, $\delta$ are the source right ascension and declination, and $\psi$ is the GW polarization angle.
The time delay $\Delta t_\mathrm{IFO}(\alpha, \delta)$ accounts for the different signal arrival times at the detectors and is also a function of the source sky location.
Throughout this paper we quote times in the Hanford frame.
So, in particular, $t_0$ refers to the ringdown start time in Hanford.
By definition, $h_+(t) = \Re\{ h(t) \}$, and $h_\times(t) = \Im \{ h(t) \}$.

\subsection{Data and Priors}
\label{sec:details}

We use the GW150914 strain data sampled at $4096\, \mathrm{Hz}$ for both the Hanford and Livingston interferometers, which was obtained from \cite{gwosc, RICHABBOTT2021100658}.
A total of $4096\,\mathrm{s}$ of data around the event was downloaded, from which the mean was subtracted (this is effectively equivalent to applying a $\sim 1\, \mathrm{Hz}$ highpass filter). 
Pre-computed power spectral densities (PSDs) associated with GW150914 from the GWTC-1 release were used \cite{gwtc1psds}. 
It has been verified our results are insensitive to the exact noise PSD used; for example, our results are unchanged when using a PSD estimated from a length of off-source data.
The analysis data consists of $4\,\mathrm{s}$ of data centered on the event GPS time ($1126259462.4\,\mathrm{s}$), and a Tukey window with an alpha parameter of 0.2 was applied to this analysis data.
The Bayesian analysis used the standard frequency-domain log-likelihood function (see, e.g., Eq.~1 in Ref.~\cite{Finch:2021qph}), with the limits of the frequency integration between $20$ and $1000\, \mathrm{Hz}$.
The choices of sampling rate and upper limit of frequency integration are discussed further in appendix \ref{app:fhigh}.

All the model parameters described in Sec.~\ref{sec:analysis} were sampled over as part of a Bayesian analysis.
For the wavelet parameters, uniform priors are used for the amplitudes $(\mathcal{A}_w \in [0,10^{-20}])$, phases $(\varphi_w \in [0,2\pi])$, frequencies $(\nu_w \in [20,200]\, \mathrm{Hz})$, and widths $(\tau_w \in [4,80]\, \tilde{M_f}$, or equivalently $\sim[1.4,27]\, \mathrm{ms})$.
Here, $\tilde{M_f}=68.779M_\odot=0.33875\,\mathrm{ms}$ is a fixed point estimate of the final, detector-frame mass (obtained using the median value from Ref.~\cite{LIGOScientific:2018mvr}) and should not be confused with the varying model parameter $M_f$.
The label-switching ambiguity among the wavelets was removed by enforcing the ordering 
$ \nu_w \leq \nu_{w+1} $ via the \emph{hypertriangulation} transformation described in Ref.~\cite{Buscicchio:2019rir}.
We sample over the wavelet central times ($\eta_w$) using a Gaussian prior in the Hanford frame with a width of $50\,\tilde{M_f}$ ($\sim 17\,\mathrm{ms}$) centered on $t_\mathrm{ref} = 1126259462.423\,\mathrm{s}$.
This choice was found to be sufficiently flexible, whilst at the same time encouraging the wavelets to accurately model the signal near the peak (see the discussion in Ref.~\cite{Finch:2021qph}).

For the ringdown, uniform priors are used for the amplitudes $(A_n \in [0,10^{-19}])$, phases $(\phi_n \in [0,2\pi])$, remnant mass $(M_f \in [40,100]\,M_\odot )$, remnant spin $(\chi_f \in [0,0.99])$ and ringdown start time $(t_0-t_\mathrm{ref} \in [-15, 15]\,\,\tilde{M_f},$ which in SI units corresponds to $\sim[-5.1,5.1]\, \mathrm{ms})$.
We use a uniform prior on $t_0$ so that the samples can be easily reweighted in post-processing (see Sec.~\ref{subsec:reweighting}). 
For the remaining parameters, we used a uniform prior over the sphere of the sky (parametrized using $\alpha$ and $\delta$) for the source location and a flat, periodic prior on the polarization angle $\psi$ in the range $0$ to $\pi$.

The nested sampling \cite{Skilling:2006gxv} algorithm as implemented in \textsc{dynesty} \cite{Speagle:2019ivv} was used to sample the posterior with 4000 live points and using the random walk sampling method with a walk length parameter of 2000.

\subsection{Reweighting} \label{subsec:reweighting}

\begin{figure}[t]
    \includegraphics[width=\columnwidth]{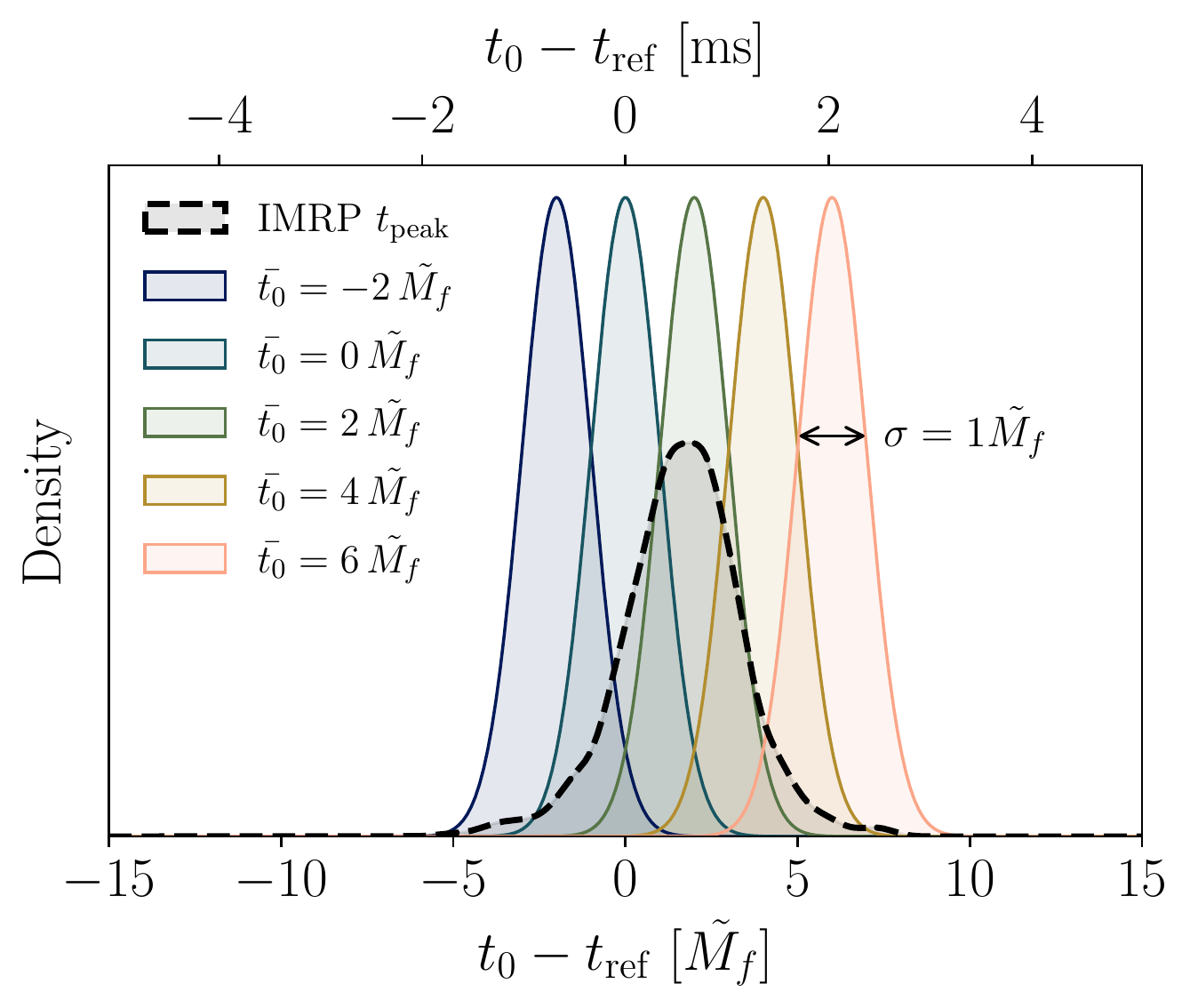}
    \caption{ \label{fig:start_time}
        Our ringdown inference is run initially using a flat, uniform prior on the ringdown start time, $t_0$, over the plot range $\pm 15 \tilde{M_f}$ relative to $t_\mathrm{ref}$ (Hanford frame).
        In post processing, the posterior samples can be reweighted to a different choice of prior on $t_0$ (see Sec.~\ref{subsec:reweighting}). 
        The different prior choices used in this paper are shown in this figure. 
        We use a sequence of narrow Gaussian priors (with different means $\bar{t_0}$ defined relative to $t_\mathrm{ref}$ and fixed standard deviation, $\sigma=1\tilde{M_f}$) as well as using the posterior on the time of peak strain from a full IMR analysis as a prior.
    }
\end{figure}

An ever present issue in ringdown analyses is the choice of ringdown start time, $t_0$, and this choice is closely related to the issue of the presence of an overtone.
To address this issue, previous time-domain analyses \cite{Isi:2019aib, Cotesta:2022pci, Isi:2022mhy} perform large numbers of Bayesian analysis runs with different choices of start time.

One key conceptual benefit of the frequency-domain approach of Ref.~\cite{Finch:2021qph} is that the ringdown start time enters as a parameter of the model and can therefore be easily marginalized over, instead of simply being fixed (although, see Ref.~\cite{Carullo:2019flw} where the ringdown start time was varied in a time-domain analysis). 
However, it is necessary to choose an informative (narrow) prior for the parameter $t_0$.

A related computational benefit of our approach is that we can do a \emph{single} Bayesian analysis run with a broad uniform prior on $t_0$. We can then explore different, narrower priors by reweighting the results in post processing. 
This is an example of \emph{importance sampling} (see, for example, \cite{RobertChristian2013MCsm}) and is the approach adopted here.
This removes the need to perform the large number of runs used to explore the effect of varying the ringdown start time when performing time-domain ringdown analyses.

Given a model that depends on parameters $\params$, a likelihood $\mathcal{L}(\mathrm{data}|\params)$, and a prior $\pi(\params)$, nested sampling can be used to draw a large number of samples $\params_i$ from the posterior, which is given by Bayes' theorem $P(\params|\mathrm{data})\propto \mathcal{L}(\mathrm{data}|\params) \pi(\params)$.
Samples from the posterior have associated weights $w_i$ (samples may often be equally weighted with $w_i=1$, but we do not require this to be the case). 
Such samples can be used to approximate integrals via a Monte-Carlo sum; $\int \mathrm{d}\params\,P(\params|\mathrm{data})f(\params)=\sum_{i}w_i f(\params_i)/W$, where $W=\sum_{i}w_i$.
If we choose a new prior $\hat{\pi}(\params)$, then the Bayesian posterior is given instead by $\hat{P}(\params|\mathrm{data})\propto \mathcal{L}(\mathrm{data}|\params) \hat{\pi}(\params)$.
We can define the new weights via
\begin{align}
    \hat{w}_i = w_i \frac{\hat{\pi}(\params_i)}{\pi(\params_i)}.
\end{align}
In this way the same samples can be used to approximate integrals of the form $\int \mathrm{d}\params\,\hat{P}(\params|\mathrm{data})f(\params)$ via the Monte-Carlo sum $\sum_{i}\hat{w}_i f(\params_i)/\hat{W}$, where $\hat{W}=\sum_{i}\hat{w}_i$.

It is also possible to reweight the Bayesian evidence for the new choice of prior.
In a GW context this approach has been used previously for inference with higher-order modes \cite{Payne:2019wmy}.
The Bayesian evidence (i.e.\ the normalization denominator in Bayes' theorem) under the original prior is given by $Z=P(\mathrm{data})=\int\mathrm{d}\params\,\mathcal{L}(\mathrm{data}|\params)\pi(\params)$.
The Bayesian evidence under the new prior, $\hat{\pi}(\params)$, is $\hat{Z}=\int\mathrm{d}\params\,\mathcal{L}(\mathrm{data}|\params)\hat{\pi}(\params)$. Using the reweighted samples to approximate the integral, it can be shown that the new evidence is given by
\begin{align}\label{eq:new_evidence}
    \hat{Z} = Z\frac{\hat{W}}{W}.
\end{align}

The process of reweighting to the new, target prior reduces the effective number of posterior samples available.
For this not to be a problem, we require the original prior to have significant support across the target prior.
Here, we reweight on just a single parameter, the ringdown start time $t_0$.
As described above, we use a uniform prior on $t_0$ as the original prior, $\pi$, in our analyses.
For the target prior we use a variety of different choices, this removes the need for performing a large number of runs with different start times. 
Our prior choices are plotted in Fig.~\ref{fig:start_time}.
Narrow Gaussians centered at different start times are used to explore the start time dependence on the results, and we use the notation $\bar{t_0}$ to indicate the mean of the Gaussian relative to $t_\mathrm{ref}$. 
For more details on the $t_0$ reweighting, see appendix \ref{app:t0_posterior_prior}.

\begin{figure*}[t]
    \captionsetup[subfigure]{labelformat=empty}
    \centering
    \;\subfloat{\includegraphics[width=.49\linewidth]{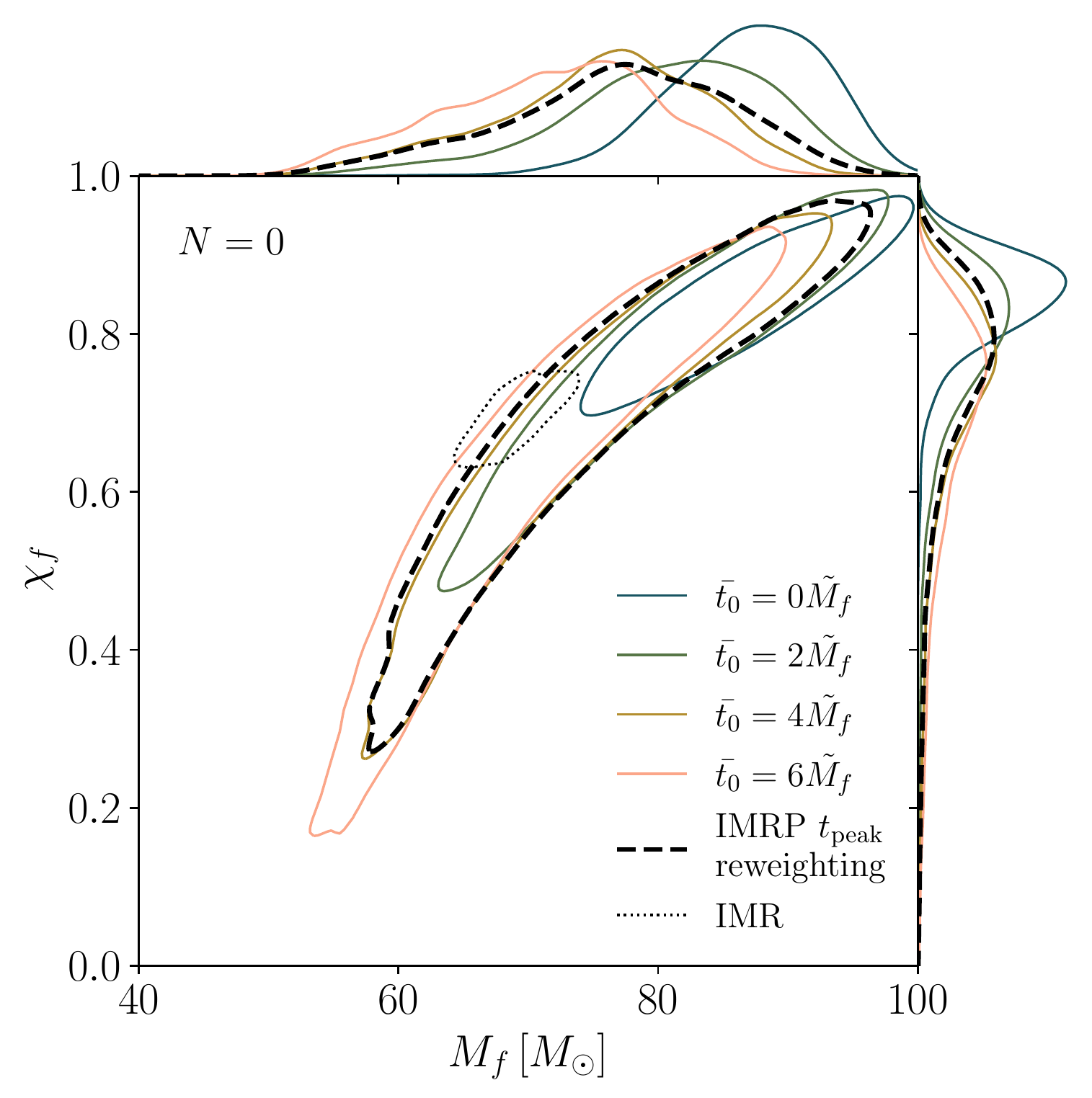}}
    \;\subfloat{\includegraphics[width=.49\linewidth]{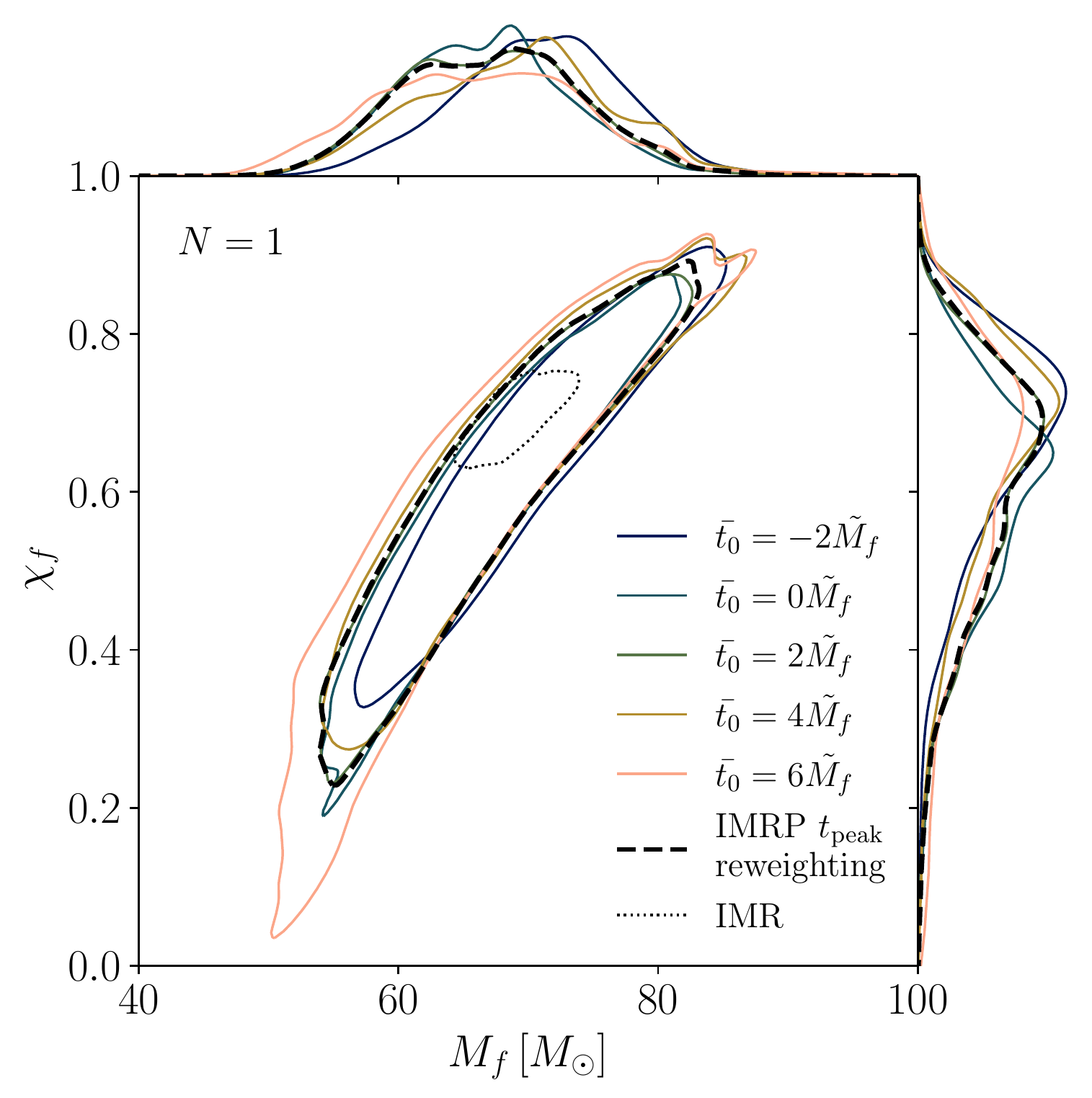}}
    \caption{ \label{fig:mass_spin_post} 
        Posterior distributions on the remnant mass, $M_f$, and dimensionless spin, $\chi_f$, for different choices of $t_0$ prior (the colors and line styles correspond to those used in Fig.~\ref{fig:start_time}). 
        \emph{Left:} the results from the $(2,2,0)$ fundamental-mode-only analysis (i.e.\ $N=0$).
        \emph{Right:} the results from the overtone analysis including the $(2,2,0)$ and $(2,2,1)$ modes (i.e.\ $N=1$).
        Each line corresponds to a different choice of $t_0$ prior. 
        Colored lines correspond to Gaussians with widths of $1 \tilde{M_f}$ and means $\bar{t_0}$ (see Fig.~\ref{fig:start_time}).
        The dashed black line corresponds to using the posterior on time of peak strain (from a full IMR analysis) as our prior, which marginalizes over uncertainty on the time of peak strain.
        Also shown for reference (dotted line) is the posterior from a full IMR analysis. 
        The main panel shows the 90\% confidence contours while the side panels show the one-dimensional marginalized posteriors.
    }
\end{figure*}

We also use the posterior on $t_\mathrm{peak}$ from a full inspiral-merger-ringdown (IMR) analysis from Ref.~\cite{Isi:2022mhy}, obtained with the \textsc{IMRPhenomPv2} (IMRP) waveform model \cite{Hannam:2013oca}, as another prior on $t_0$. 
Our aim in doing this is to marginalize over our uncertainty on the ringdown start time, $t_0$. 
We emphasize that this is achieved here by using the posterior on the time of peak strain as a prior on $t_0$; this is motivated by the observations of Refs.~\cite{Giesler:2019uxc, Bhagwat:2019dtm, Ota:2019bzl, Cook:2020otn, JimenezForteza:2020cve, Dhani:2020nik, Finch:2021iip, Forteza:2021wfq, Dhani:2021vac, MaganaZertuche:2021syq} described above, which show that generically the ringdown can be considered to start at around this time.

\section{Results}\label{sec:results}

There are several ways to investigate and quantify the evidence for additional QNMs in the ringdown.
Sec.~\ref{subsec:overtone} contains the results of a series of analyses designed to study the presence of a possible overtone in GW150914.
Sec.~\ref{subsec:verify} contains the results of a series of analyses designed to test whether or not what has been detected really is an overtone and is not the accumulation of other effects.
Sec.~\ref{subsec:noise} describes further checks on the stability of the results, and
Sec.~\ref{subsec:other_results} contains some additional results that further demonstrate the capabilities of the frequency-domain approach to ringdown analysis.

Throughout Secs.~\ref{subsec:overtone} and \ref{subsec:verify}, we compare our results with those in Refs.~\cite{Cotesta:2022pci} and \cite{Isi:2022mhy}. 
This is done in the hope of helping to resolve the controversy over the evidence for a ringdown overtone in GW150914. 
However, it should be stressed that our results are produced using a very different method and care should therefore be taken in making direct comparisons.
Although the frequency-domain analysis is formally equivalent to the time-domain analysis in a particular limit (as discussed in the introduction, and in more detail in Ref.~\cite{Finch:2021qph}) we do not take this limit in a practical analysis. Furthermore, the frequency-domain analysis is further generalized with respect to the time-domain analysis in that it marginalizes over parameters such as the sky position and ringdown start time (which are fixed in the analyses of Refs.~\cite{Cotesta:2022pci, Isi:2022mhy}).
Results from our frequency-domain analyses should therefore not be expected to agree perfectly with those from previous time-domain analyses.

\begin{figure*}[t]
    \centering
    \includegraphics[width=1.9\columnwidth]{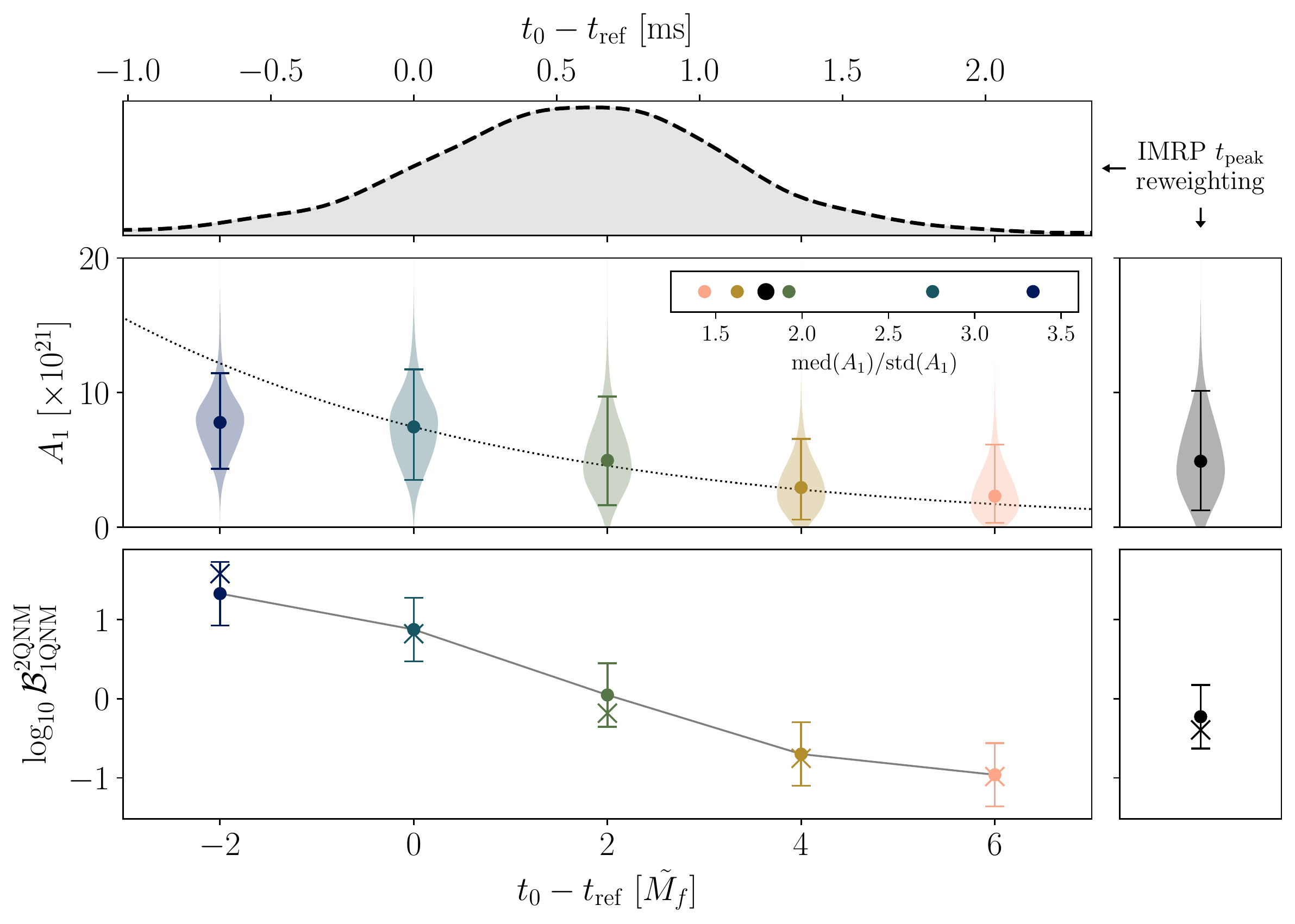}
    \caption{ \label{fig:overtone_amplitude}
        Posteriors on the overtone amplitude, and Bayes' factors in favor of the overtone model for different choices of $t_0$ prior (the colors and line styles correspond to those used in Fig.~\ref{fig:start_time}).
        \emph{Top:} posterior on the time of peak strain in the Hanford frame, from a \textsc{IMRPhenomPv2} analysis, as in Fig.~\ref{fig:start_time} (and originally from Ref.~\cite{Isi:2022mhy}). 
        \emph{Middle:} overtone amplitude posteriors for different choices of $t_0$ prior. The left panel corresponds to Gaussian priors with standard deviation $1\tilde{M_f}$, centered at the time they are plotted.
        The dotted line indicates the expected exponential decay of the $A_1$ mode; this is included merely to guide the eye and was produced using the median mass and spin values from the full IMR analysis and the median value of $A_1$ from the $\bar{t_0} = 0$ prior.
        The right panel corresponds to using the \textsc{IMRPhenomPv2} time of peak strain as a prior.
        For earlier start times the posteriors on the amplitude are peaked further away from zero; this is quantified in the inset plot where the ratio of the median to the standard deviation of the $A_1$ posterior is plotted.
        \emph{Bottom:} the Bayes' factor in favor of the overtone model for each prior choice; circles with error bars show the Bayes' factor calculated from nested sampling (with errors estimated by the sampler) while the crosses show the results calculated using the Savage-Dickey density ratio.    
    }
\end{figure*}

\subsection{Presence of an overtone}\label{subsec:overtone}

In order to investigate the presence of an overtone in the GW150914 ringdown, we initially perform two analyses using the model described in Sec.~\ref{sec:model}: one analysis uses only the fundamental QNM ($N=0$) and the other includes the first overtone ($N=1$).
Aside from the inclusion of the overtone in the ringdown (which introduces two additional parameters: an amplitude and a phase), these two analyses are otherwise identical.

In Fig.~\ref{fig:mass_spin_post} we plot the posterior distributions on the remnant BH mass, $M_f$, and dimensionless spin, $\chi_f$, for both of these analyses.
Results are shown for the different choices of the prior on the ringdown start time shown in Fig.~\ref{fig:start_time} (these results were obtained by reweighting the samples obtained with a flat prior using the approach described in Sec.~\ref{subsec:reweighting}).
The earliest start time ($\bar{t_0}=-2\tilde{M_f}$) is omitted from the fundamental-only ($N=0$) plot in the left-hand panel of Fig.~\ref{fig:mass_spin_post} because of a low number of posterior samples at these time (see appendix \ref{app:t0_posterior_prior}).
Also shown for comparison are the much tighter constraints resulting from the full IMR analysis.
These IMR posterior samples were obtained from Ref.~\cite{maximiliano_isi_2022_5965773}, which (as detailed in Refs.~\cite{Isi:2019aib,Isi:2022mhy}) are obtained from applying fitting formulas to the samples available at Ref.~\cite{gwtc1samples}. 
When only the fundamental QNM is used ($N=0$), and when the analysis is started at early times (e.g.\ $t_0 - t_\mathrm{ref}\lesssim -2\tilde{M_f}$) our posteriors on the remnant parameters are biased towards high values of $M_f$ and $\chi_f$.
This behavior is expected; a single QNM is only able to model the ringdown signal starting well after the time of peak strain.
Including an overtone ($N=1$) allows the ringdown analysis to start at earlier times, as can be seen by the removal of the bias in the right panel. 
This improvement is suggestive that the data supports the inclusion of an overtone.
Additionally, using an earlier ringdown start time increases the SNR in the ringdown and reduces the posterior width; this effect can be seen in both the $N=0$ and $N=1$ analyses.

Our results in Fig.~\ref{fig:mass_spin_post} can be compared to the corresponding results of the time-domain analyses shown in Fig.~1 from Cotesta et al. \cite{Cotesta:2022pci} and Figs.~4 and 5 from Isi \& Farr \cite{Isi:2022mhy}.
In general terms, there is broad agreement between all three sets of results. 
In particular, all three sets of authors find that the overtone analyses ($N=1$) always gives results that are more consistent with the IMR result and get increasingly broader for later choices of the ringdown start time.
All sets of authors also find that for the fundamental-only analysis ($N=0$) starting at early times (i.e.\ $t_0  - t_\mathrm{ref}\lesssim 0$) leads to posteriors that are inconsistent with the IMR result.
However, there are subtle differences between the various results.
Our results with $N=0$ and early start times gives posteriors biased to large values of $M_f$ and $\chi_f$; this is also seen in Ref.~\cite{Isi:2022mhy}, but not in Ref.~\cite{Cotesta:2022pci} (where the posterior consistently reaches lower values of $\chi_f$).
Our results with $N=0$ and late start times (i.e.\ $t_0 - t_\mathrm{ref}\gtrsim 4\tilde{M_f}$) are partially consistent with the IMR results; this is also seen in Ref.~\cite{Cotesta:2022pci}, but not in Ref.~\cite{Isi:2022mhy} who never find consistency with the IMR result for any choice of start time.
Finally, when including the overtone ($N=1$) and starting at late times, Ref.~\cite{Cotesta:2022pci} find results that are consistent with $\chi_f=0$ (i.e.\ a Schwarzschild BH) at 90\% confidence, in stark disagreement with Ref.~\cite{Isi:2022mhy} who find $\chi_f\gtrsim 0.2$. 
Our results are in better agreement with those of Ref.~\cite{Isi:2022mhy}.

In the middle panel of Fig.~\ref{fig:overtone_amplitude} we investigate our $N=1$ overtone analysis further by plotting the one-dimensional marginalized posteriors for the amplitude, $A_1$, of the QNM overtone.
An amplitude posterior peaked away from zero has been suggested (particularly by Ref.~\cite{Isi:2019aib}) as one good indication for the presence of an overtone in the data.
As expected, the QNM overtone decays quickly and when starting at later times we find a small value for the amplitude.
The degree to which the $A_1$ posterior is peaked away from zero can be quantified using the ratio between the median and standard deviation; this is plotted in the inset of the middle panel of Fig.~\ref{fig:overtone_amplitude}.
For values of $\bar{t_0}$ between $-2\tilde{M_f}$ and $+6\tilde{M_f}$, we find posteriors on $A_1$ that are peaked away from zero at between $1.44$ and $3.34\sigma$.
If we reweight using the IMRP $t_{\rm peak}$ prior, we find a posterior peaked away from zero at $1.79\sigma$.

\begin{figure}[t]
    \includegraphics[width=0.49\textwidth]{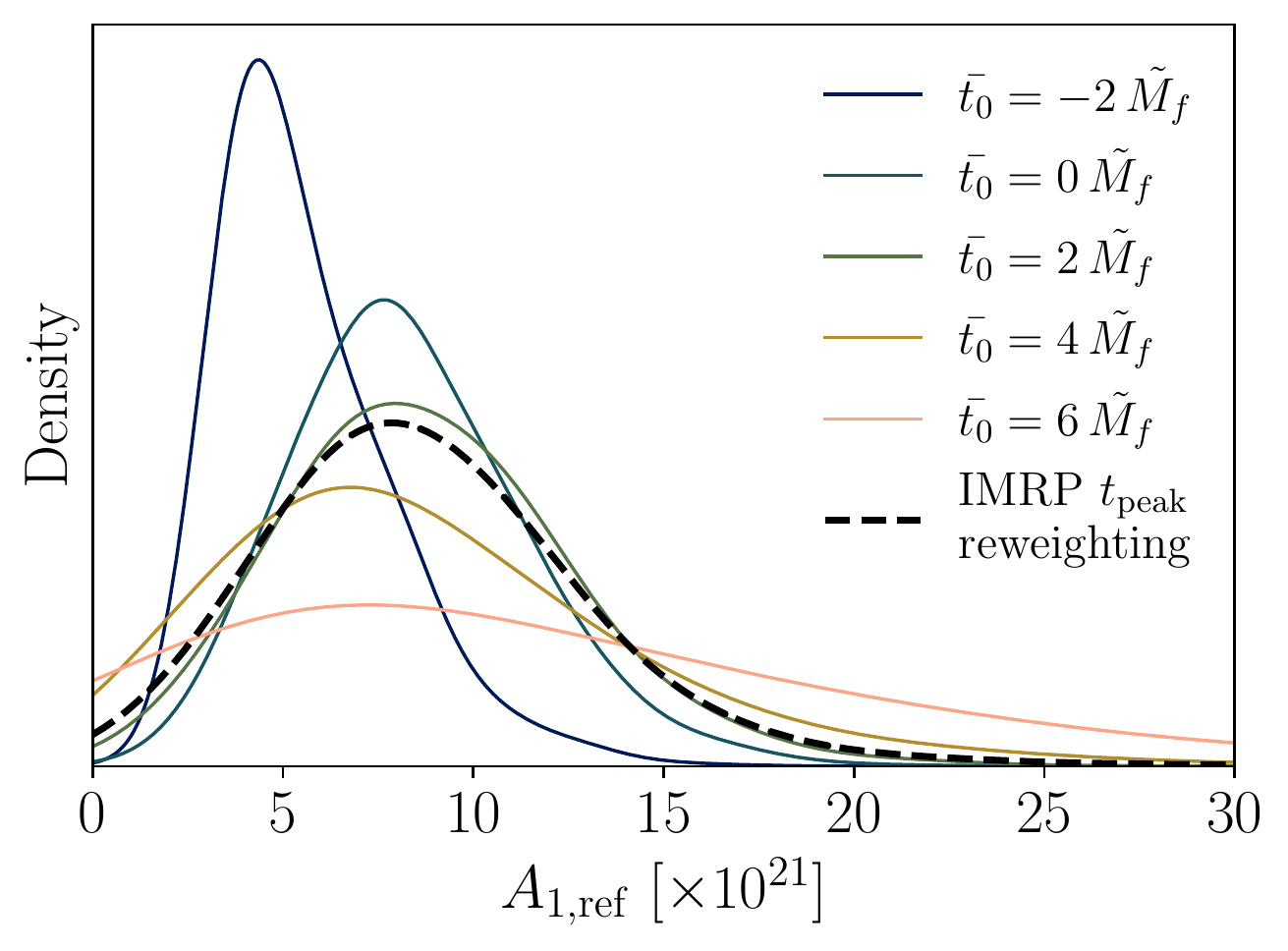}
    \caption{ \label{fig:amp_at_tref}
        Posteriors on the overtone amplitude from our $N=1$ overtone analysis, rescaled to a fixed reference time of $t_{\rm ref}$.
        The rescaling does not significantly affect the significance with which the posteriors are peaked away from zero.
        The colors and line styles indicate the prior used on $t_0$ and correspond to those used in Fig.~\ref{fig:start_time}.
    }
\end{figure}

Our results in the middle panel of Fig.~\ref{fig:overtone_amplitude} can be compared to the corresponding results of the time-domain analyses shown in Fig.~1 of Ref.~\cite{Isi:2022mhy} and Fig.~2 of Ref.~\cite{Cotesta:2022pci}.
All three sets of authors find values of $A_1$ that are smaller at later times, consistent with the expected exponential decay of the overtone, but they disagree on the absolute value of the amplitude and the significance with which a zero amplitude can be excluded.
Refs.~\cite{Isi:2019aib, Isi:2022mhy} find the largest values; they report a posterior peaked $3.6\sigma$ away from zero.
Ref.~\cite{Cotesta:2022pci} finds much smaller values which are consistent with zero for many choices of start time.
These analyses use essentially the same method and should therefore agree exactly.
Our result, produced using a different method, lies somewhere in between; we do find nonzero values are preferred for a range of start times, but only with a modest significance of $\sim 1.79\sigma$ for our preferred IMRP $t_{\rm peak}$ prior which we consider to be the best description of our uncertainty on the ringdown start time.

The comparison of our results with those of Refs.~\cite{Isi:2019aib, Cotesta:2022pci, Isi:2022mhy} is complicated by the fact that we use subtly different definitions for the amplitude. 
The time-domain analyses naturally define the mode amplitudes at a fixed time, usually $t_0$.
Our frequency-domain analysis also defines the mode amplitudes at $t_0$, but this start time is then varied as part of the analysis, blurring the exact time at which the amplitude is defined.
This is a fairly small effect for the narrow Gaussian priors, but more significant for the wider IMRP $t_{\rm peak}$ prior.
We can correct for this effect by rescaling all the overtone amplitudes to any fixed reference time (here we use $t_{\rm ref}$) using the known decay rate for the QNMs;
\begin{align}
    A_{1,\mathrm{ref}} = A_1 \exp\left(\frac{t_0-t_{\mathrm{ref}}}{\tau_{221}(M_f,\chi_f)}\right),
\end{align}
where $\tau_{221}(M_f, \chi_f)$ is the exponential decay time of the $(2,2,1)$ QNM and is a function of the remnant mass and spin.
This rescaling can be done for any QNM and the resulting amplitude parameters $A_{\ell m n,\mathrm{ref}}$ are more directly comparable with the amplitudes used in time-domain analyses.
Posteriors on $A_{1,\mathrm{ref}}$ are shown in  
Fig.~\ref{fig:amp_at_tref}.

\begin{figure}[t]
    \includegraphics[width=0.49\textwidth]{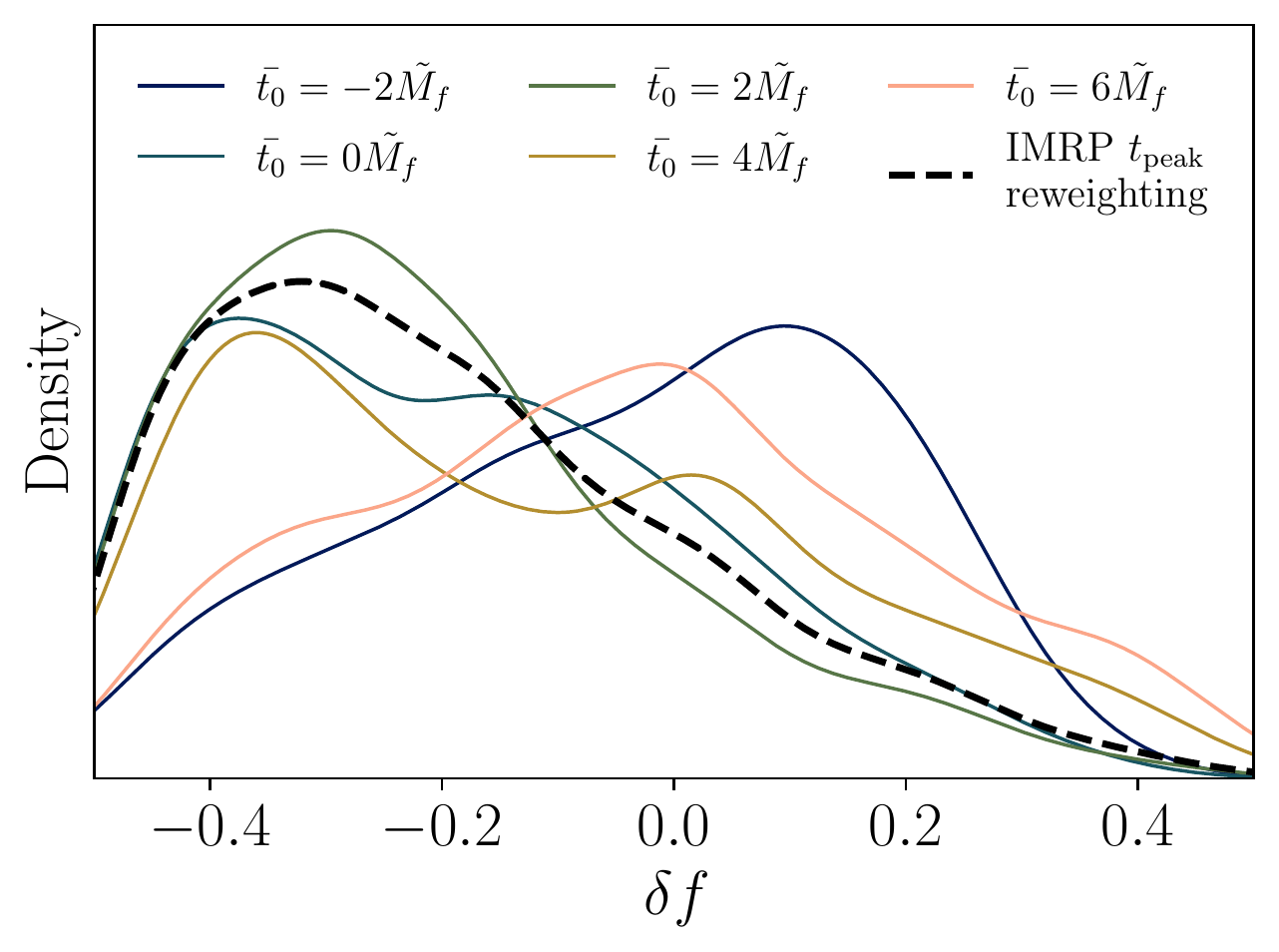}
    \caption{ \label{fig:delta_f}
        Posteriors on the deviation parameter from the Kerr value for the real part of the overtone frequency.
        The colors and line styles distinguish different choices for the $t_0$ prior and correspond to those used in Fig.~\ref{fig:start_time}.
        The mode frequency is given by $f_{221}^{\rm Kerr} \exp(\delta f)$, so that $\delta_f=0$ is the expected result for the Kerr metric.
        For all choices of $t_0$ prior the data is consistent with $\delta f=0$.
    }
\end{figure}

In the bottom panel of Fig.~\ref{fig:overtone_amplitude} we plot the Bayes' factors between the fundamental only ($N=0$) and overtone ($N=1$) analyses.
This is defined as $\mathcal{B}_{\rm 1QNM}^{\rm 2QNM}=Z_{N=1}/Z_{N=0}$.
The Bayes' factor has been suggested (particularly by Ref.~\cite{Cotesta:2022pci}) as another good way for quantifying the support for an overtone in the data.
The Bayes' factor was computed in two different ways.
Firstly, \textsc{dynesty} was used to calculate the evidences $Z_{N=0}$ and $Z_{N=1}$ for both of the analyses described above, and these were reweighted to the desired $t_0$ prior using Eq.~\ref{eq:new_evidence}. 
Nested sampling also returns an estimate for the error on the evidences, and these are used to plot the error bars in Fig.~\ref{fig:overtone_amplitude}.
Secondly, exploiting the fact that the $N=0$ model is nested within the $N=1$ model, the Bayes' factors were computed using the posterior on $A_1$ from the $N=1$ analysis to find the Savage-Dickey density ratio~\cite{Dickey:1971}. 

Our results in the bottom panel of Fig.~\ref{fig:overtone_amplitude} can be compared to the corresponding results of the time-domain analyses shown in Fig.~7 of Ref.~\cite{Isi:2022mhy} and Fig.~2 of Ref.~\cite{Cotesta:2022pci}.
Ref.~\cite{Cotesta:2022pci} computes the Bayes' factors using the ratio of evidences evaluated with nested sampling, whereas Ref.~\cite{Isi:2022mhy} computes Bayes' factors using Savage-Dickey density ratios.
All sets of authors find Bayes' factors that decrease for later ringdown start times, although they disagree on the exact value.
Ref.~\cite{Isi:2022mhy} finds the strongest log-evidence of $\sim 1.7$ at $t_0-t_{\rm ref} \sim 0$.
Ref.~\cite{Cotesta:2022pci} finds slightly \emph{negative} log-evidence starting at this time.
Again, our result lies somewhere in between, we find a moderate log-evidence of $\sim 1.0$ when marginalizing over a narrow prior on $t_0$ centered at this time.
If we instead marginalize over the time of peak strain using the broader IMRP $t_{\rm peak}$ prior, the evidence is slightly negative.
However, as discussed in Sec.~\ref{sec:discussion} below, we consider the actual values of the Bayes factors to be less important than their trend with varying start time.

\begin{figure}[b]
    \includegraphics[width=0.49\textwidth]{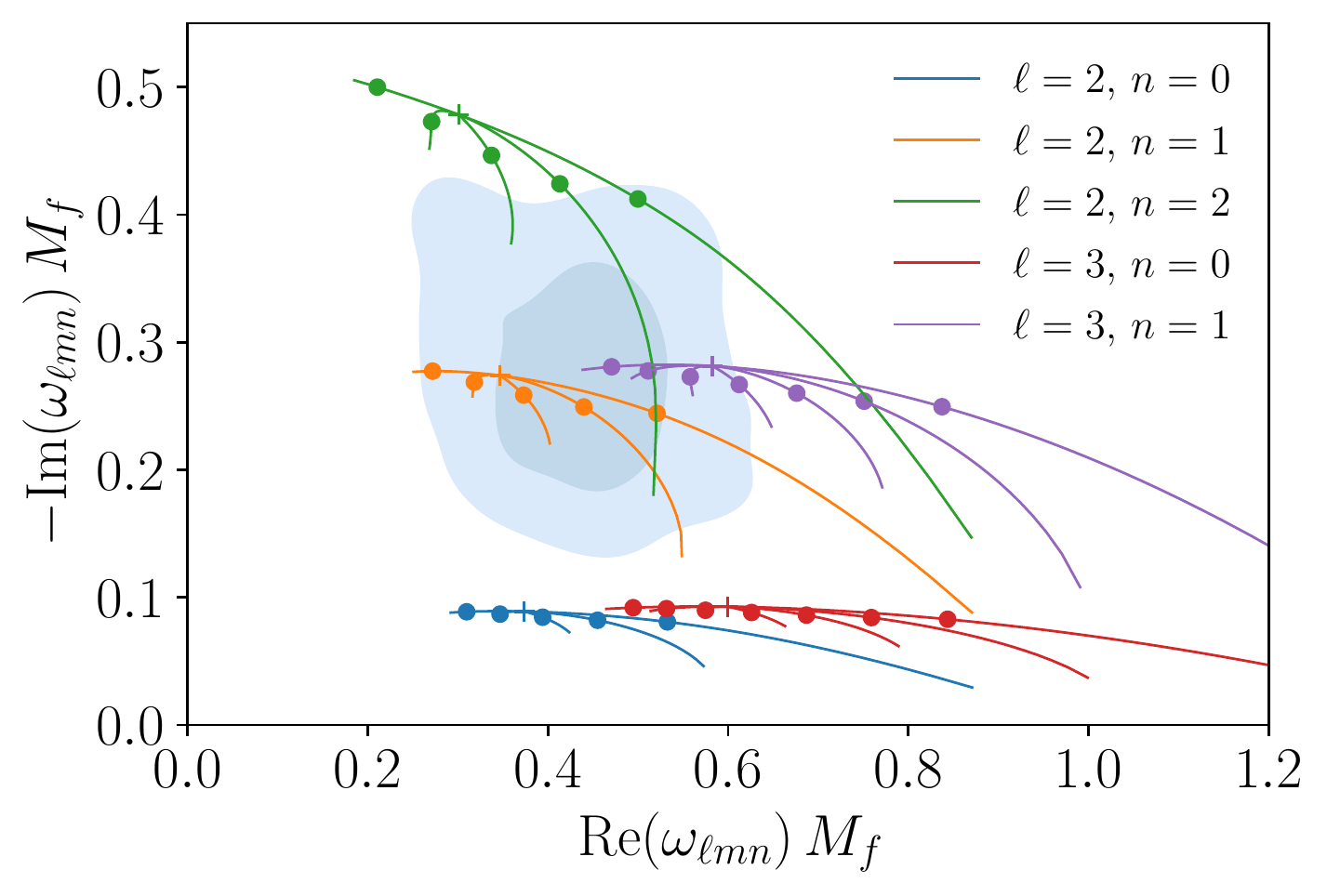}
    \caption{ \label{fig:other_QNMs}
        The posterior on the dimensionless complex frequency of the second QNM (50\% and 90\% regions), assuming the first is the fundamental $(\ell,|m|,n)=(2,2,0)$ mode.
        Lines indicate the Kerr frequencies parametrized by the remnant spin; dots and crosses indicate points with $\chi_f=0.7$ and $0$ respectively.
        Lines are colored according to their $\ell$ and $n$ indices and the $m$ index increases left to right in each set.
        The frequency of the second QNM is consistent with the expected $(2,2,1)$ overtone, but also with several other modes.
        However, all fundamental modes (those with $n=0$) are excluded.
    }
\end{figure}

\subsection{The nature of the overtone}\label{subsec:verify}

The results of the previous section show that there is tentative evidence for something beyond the fundamental $(2,2,0)$ mode in the GW150914 data. 
In the previous section it was assumed that this is the $(2,2,1)$ QNM overtone; this is motivated by our expectations from numerical relativity experiments (see, for example, Ref.~\cite{Giesler:2019uxc}). 
In this section, we address this assumption by measuring the frequency and amplitude of the QNM overtone and comparing with the expectations from GR.

Fig.~\ref{fig:delta_f} shows the results of a third ringdown analysis that also includes two QNMs.
In this analysis the complex frequency of the second QNM is allowed to deviate from the Kerr overtone value. 
This differs from the $N=1$ overtone analysis described above, where the frequency of the overtone was fixed by the remnant mass and spin to the Kerr value, $\omega_{221} = 2\pi f_{221} - i/\tau_{221}$.
Recovering a value of $\delta f$ consistent with zero has been suggested (particularly by Ref.~\cite{Isi:2022mhy}) as further evidence for the presence of an overtone; otherwise, it might be expected that the extra parameters would fit to the noise and would not recover the Kerr value.
We use the parametrization from Ref.~\cite{Isi:2022mhy}; the complex frequency of the second QNM is now $\omega_{221} = 2\pi f-i/\tau$, where $f=f_{221}\exp(\delta f)$ and $\tau=\tau_{221}\exp(\delta \tau)$. 
This introduces the two new dimensionless parameters $\delta f$ and $\delta \tau$ into the model, for which we use uniform priors in the range $[-0.5,\, 0.5]$.
The $\delta \tau$ parameter is not well constrained, therefore we focus initially on $\delta f$.
We find posteriors on $\delta f$ consistent with zero for all choices of $t_0$ prior with standard deviations $\sim 0.2$. 
This is consistent with what was found in Ref.~\cite{Isi:2019aib} and can be viewed as a test of the no-hair theorem at the $\sim 20\%$ level.

Our results in Fig.~\ref{fig:delta_f} can be compared with Fig.~2 of Ref.~\cite{Isi:2022mhy}. 
Our preferred run, using the IMRP $t_{\rm peak}$ prior on $t_0$, is broadly consistent with that result.
However, what is notable about our results is that we do not find a significant broadening of the posterior for later choices of the start time. 
This was found by Ref.~\cite{Isi:2022mhy} and would be expected if an overtone was present, as both the overtone amplitude and ringdown SNR decaying with later ringdown start times.

To investigate this further, we use the results of the ringdown analysis where the frequency of the overtone is allowed to vary freely to address another important question. 
If the data does indeed contain a second QNM, can we determine which mode it is?
Theoretical studies of numerical relativity simulations suggest that the $(\ell,|m|,n)=(2,2,1)$ will be the next most prominent, especially for early start times \cite{Giesler:2019uxc}. 
In Fig.~\ref{fig:other_QNMs} we plot the posterior on the \emph{dimensionless} complex frequency (allowing both the real and imaginary parts to vary freely) of the second QNM, $\omega M_f$.
This plot uses the value for $M_f$ calculated from the complex frequency of the first QNM, assuming this is the expected $(2,2,0)$ fundamental mode of Kerr.
We find that we can confidently conclude that the second mode is an overtone ($n\geq 1$) but that it is not possible to say from the data alone exactly which overtone. 
For example, the modes $(2,2,1)$ and $(2,1,1)$ are both equally compatible with the data. 
In general, when searching for additional QNMs it is necessary to be guided by our prior expectations regarding which modes are expected to be excited with the highest amplitudes.

\begin{figure}[t]
    \includegraphics[width=0.49\textwidth]{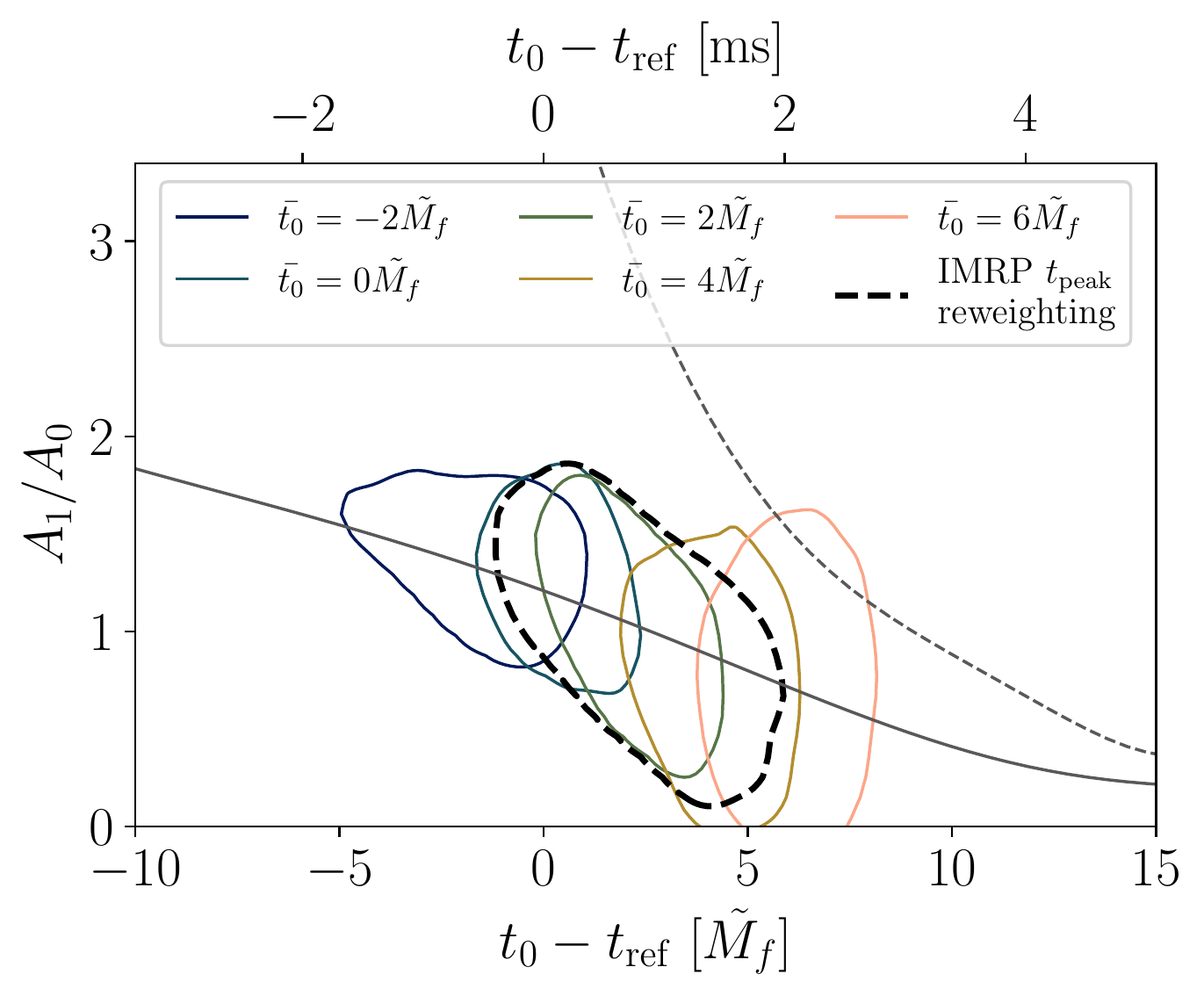}
    \caption{ \label{fig:amp_ratio}
        Posteriors on the amplitude ratio $A_1/A_0$ from our $N=1$ overtone analysis. 
        The 90\% contours are plotted, with the colors and line styles indicating the $t_0$ prior and correspond to those used in Fig.~\ref{fig:start_time}.
        The solid gray curve shows the results of a two-QNM fit to the numerical relativity simulation SXS:BBH:0305 which has parameters consistent with GW150914.
        The dashed gray curve shows the results of a multi-QNM fit to SXS:BBH:0305 which follows closely the expected exponential decay rate for the amplitude ratio.
    }
\end{figure}

We now turn our attention to the measured amplitude $A_1$ and whether this matches the theoretical expectations for the $(2,2,1)$ overtone. 
For convenience, we choose to work with the amplitude ratio $A_1/A_0$ which eliminates factors common to all modes, such as the distance to the source. 
Because the two QNMs decay exponentially at different rates, the amplitude ratio depends strongly on the chosen ringdown start time.
Our two-dimensional posteriors on the amplitude ratio and ringdown start time are plotted in Fig.~\ref{fig:amp_ratio}.
As expected we find that the amplitude ratio decreases for later start times, and the error on the amplitude ratio increases for later start times because of the decreasing SNR in the ringdown.

In order to check whether this is consistent with the theoretical expectation for an overtone we compare with fits to the numerical relativity simulation SXS:BBH:0305 \cite{2016CQGra..33x4002L, sxs_catalog} which has parameters consistent with GW150914.
Fixing the remnant mass and spin to the values reported in the simulation metadata, we perform QNM least-squares fits to this simulation for a range of ringdown start times using the code previously developed in Ref.~\cite{Finch:2021iip}.
Results are shown in Fig.~\ref{fig:amp_ratio} for two such fits. 
Firstly, we performed a two-QNM fit intended to mimic the analysis of the real GW150914 data described in Sec.~\ref{subsec:overtone} above. 
In this analysis the ${}_{-2}Y_{22}$ spherical harmonic mode of the simulation is modeled as a sum of the $(2,2,0)$ and $(2,2,1)$ QNMs and the amplitude ratio is recorded. 
The results from this two-QNM fit agree very well with what is seen in the real data giving us further confidence that there is nothing unexpected present in the data and that our results are not unduly affected by noise fluctuations (see also the discussion in Sec.~\ref{subsec:noise}). 
Secondly, we perform a full multi-QNM fit to all the spherical harmonic modes (up to and including $\ell = 8$) with a ringdown model that includes all QNMs (including both prograde and retrograde modes) up to $\ell = 8$ and $n = 7$ (1232 QNMs in total).
The ratio of the amplitudes of the $(2,2,0)$ and $(2,2,1)$ prograde modes from this fit behaves very differently; the ratio follows very closely a exponential time evolution which can be understood in terms of the difference between the two QNM decay times.

The fact that the two-QNM analysis gives a very different amplitude ratio compared to the full multi-QNM analysis for ringdown start times near the peak strain is related to the extreme destructive interference observed in the QNM overtone fits of Refs.~\cite{Giesler:2019uxc, Bhagwat:2019dtm, Ota:2019bzl, Cook:2020otn, JimenezForteza:2020cve, Dhani:2020nik, Finch:2021iip, Forteza:2021wfq, Dhani:2021vac, MaganaZertuche:2021syq} with large values of $N$.
This shows that the amplitude $A_1$ recovered from a two-QNM analysis is not purely the amplitude of the first overtone but also includes significant contributions from higher overtones and other harmonics. 
However, absorbing these contributions into the first overtone introduces a systematic bias in the remnant properties that is smaller than the statistical uncertainty; this can be seen in, for example, Fig.~\ref{fig:mass_spin_post} and Sec. IV\,C of Ref.~\cite{Giesler:2019uxc}. 
For this reason, it still makes sense to describe the results of the two-QNM analysis as a measurement of the overtone, even though there are undoubtedly other contributions present in the signal.

\subsection{The effect of noise and sampling rates}\label{subsec:noise}

One of the key claims made in Ref.~\cite{Cotesta:2022pci} was the overtone detection was highly sensitive to noise fluctuations.
This was disputed by Ref.~\cite{Isi:2022mhy}.
In order to address this issue, we performed a noise injection study mirroring closely what was done in Ref.~\cite{Cotesta:2022pci}.
The results of this injection study are presented in appendix \ref{app:inj}.
As expected, the results of injecting into different noise realizations show some scatter.
However, this scatter is not larger than expected and we are unable to reproduce the claim in Ref.~\cite{Cotesta:2022pci} with our (very different) analysis method. 

It has been suggested \cite{WillMaxTGRtelecon} that the results of ringdown analyses, in particular those including highly damped overtones that contain significant power at high frequencies, might be sensitive to aliasing effects when using downsampled strain data.
This is discussed in more detail in appendix \ref{app:fhigh} where it is argued that our results are insensitive to changes in both the sampling rate of the data and the value of $f_{\rm high}$.

\subsection{Other results} \label{subsec:other_results}

\begin{figure}[t]
    \includegraphics[width=0.49\textwidth]{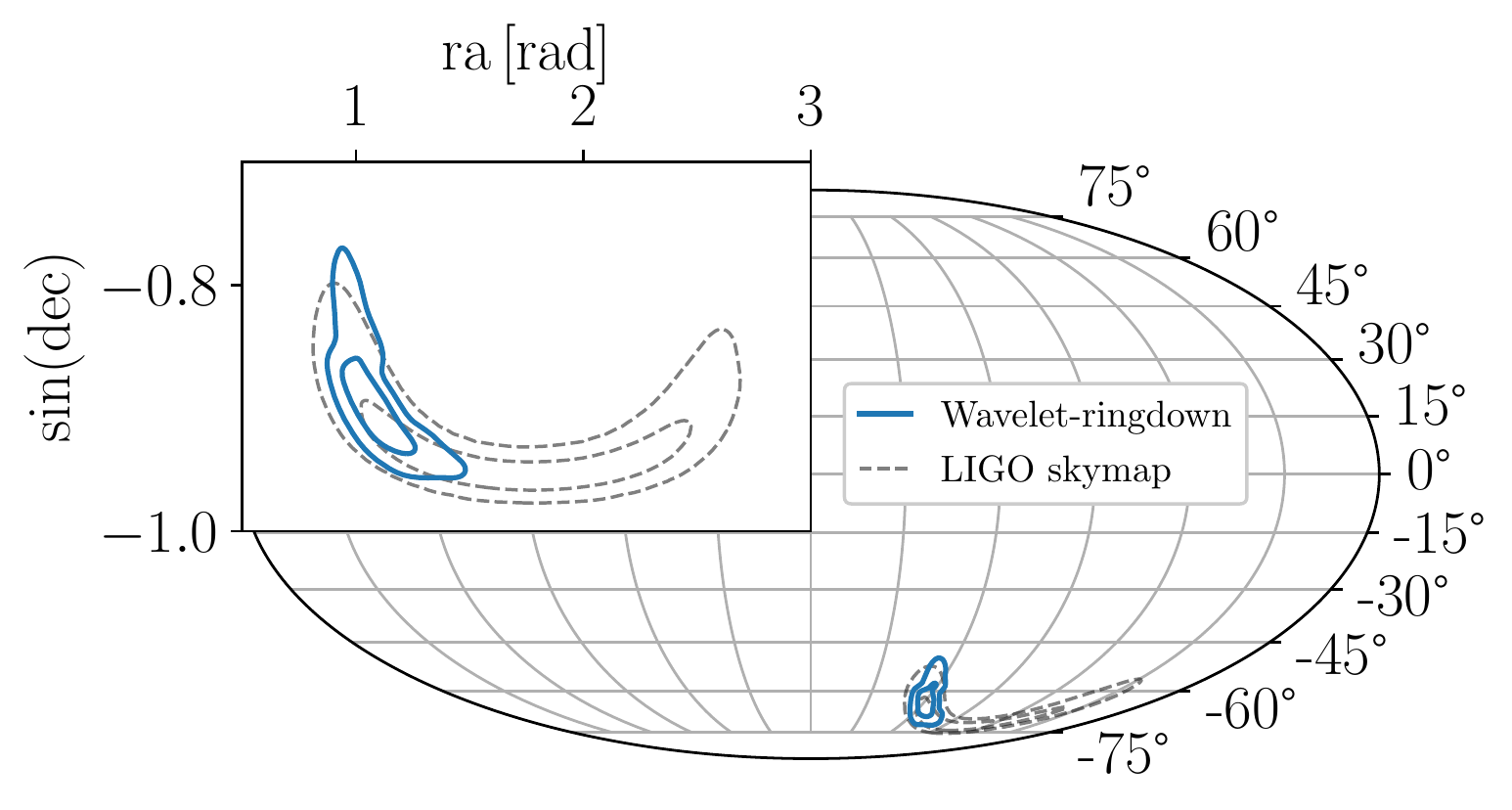}
    \caption{ \label{fig:skymap}
        Posterior on the source sky position using geocentric coordinates in Mollweide projection.
        Shown in blue is the results from the $N=1$ overtone analysis using the IMRP $t_{\rm peak}$ reweighting for the prior on the ringdown start time.
        The LIGO skymap for this event is shown by the dashed black line for comparison.
        The inset plot shows a zoomed-in map plotted using right ascension and the sine of the declination.
        In both cases, 50\% and 90\% contours are plotted.
    }
\end{figure}

One important benefit of the frequency-domain approach to ringdown data analysis introduced in Ref.~\cite{Finch:2021qph} and used here is that it naturally allows us to search (and hence to numerically marginalize) over source sky position and ringdown start time. 
This should be contrasted with the treatment of these parameters in most time-domain analyses where these parameters are fixed, potentially biasing the results. (Although it is technically possible to search over the sky in a time-domain analysis \cite{Carullo:2019flw, Isi:2021iql}, this is rarely done in practice.)
To emphasize this, we plot the posterior on the sky location of GW150914 from our $N=1$ overtone analysis reweighted to the IMRP $t_{\rm peak}$ prior on the ringdown start time.
This can be compared with the publicly available LIGO sky posterior for GW150914 obtained using the samples from \cite{skysamples}.
This is shown in Fig.~\ref{fig:skymap}.
As discussed in \cite{Finch:2021qph}, it should be emphasized that this sky posterior is not a ringdown-only result because much of the information is also coming from the wavelets used to model the inspiral-merger portion of the signal.

\begin{figure*}[t]
    \centering
    \includegraphics[width=2\columnwidth]{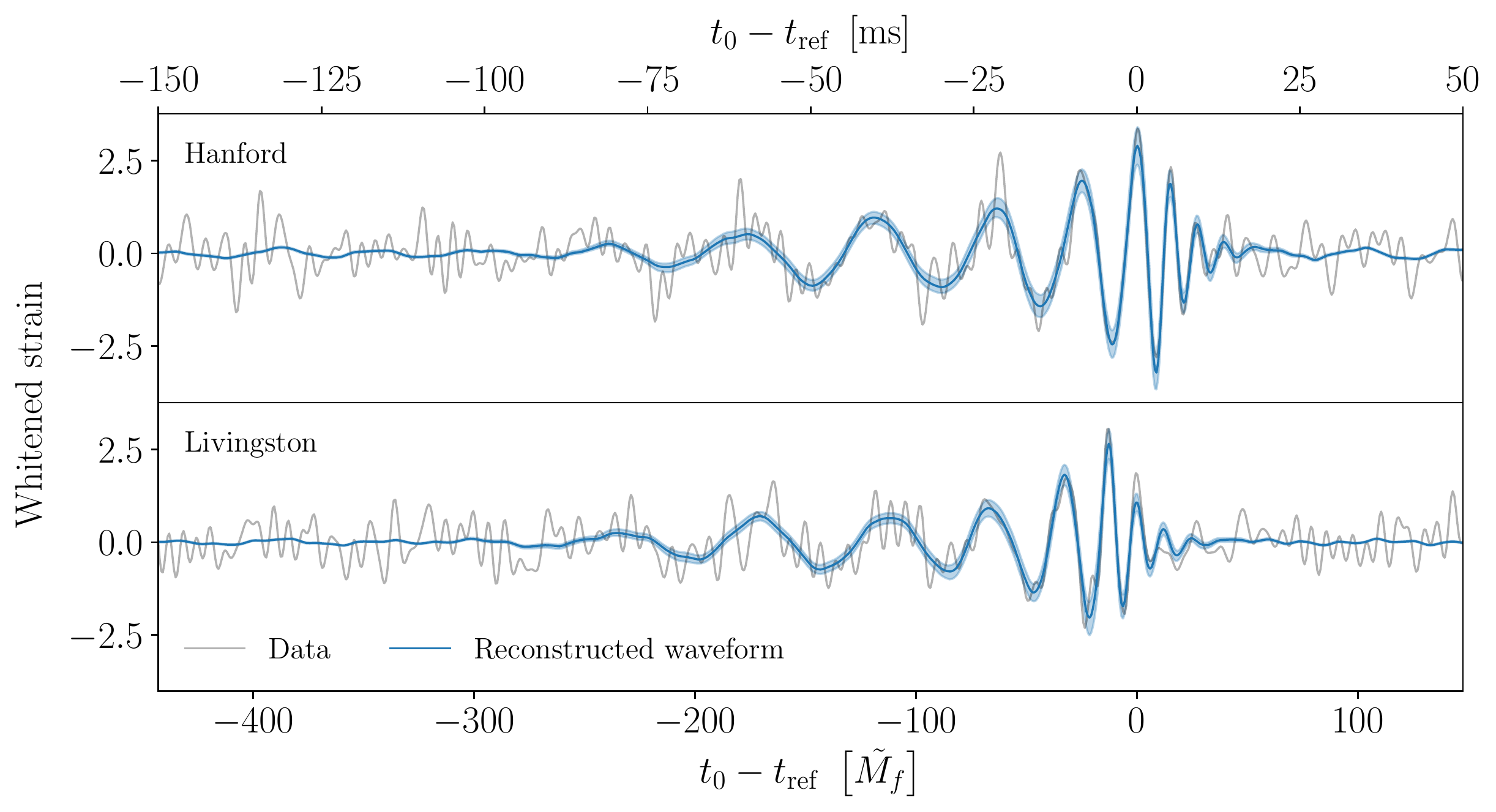}
    \caption{ \label{fig:waveform}
        Posterior on the reconstructed whitened waveform.
        Shown in gray is the strain data from both LIGO interferometers (\emph{top}: Hanford, \emph{bottom}: Livingston) whitened according to the noise amplitude spectral density in the detector and bandpass filtered between 32 and $512\,\mathrm{Hz}$ for clarity.
        Shown in blue is the waveform reconstruction from the $N=1$ overtone analysis with the IMRP $t_{\rm peak}$ reweighting for the prior on the ringdown start time.
        The blue lines and shaded regions indicate median and the 90\% credible interval. 
        The signal is plotted as a function of time from $t_{\rm ref}$ using both SI and natural units on the upper and lower $x$-axis respectively.
    }
\end{figure*}

Because the inspiral and merger parts of the signal are being modeled using truncated wavelets as part of the frequency-domain ringdown analysis, this allows us to plot a full waveform reconstruction from our results.
This reconstruction is shown in Fig.~\ref{fig:waveform} for our $N=1$ overtone analysis reweighted to the IMRP $t_{\rm peak}$ prior.
The full waveform model used in our analysis is discontinuous at $t_0$. However, as discussed in Ref.~\cite{Finch:2021qph}, the whitened waveform reconstruction plotted here is smooth; this is a result of marginalizing over the location of the discontinuity at $t_0$, the waveform model ``learning'' the continuity from the data, and the whitening process used to make the figure.
This waveform reconstruction uses the posterior on all of the model parameters, including those for the wavelets; more details on these parameters are given in appendix \ref{app:W3}.

\section{Discussion and Conclusions}\label{sec:discussion}

The main motivation for this work comes from the ongoing discussion in the literature about whether a ringdown overtone can be confidently detected in the GW150914 data. 
In particular, the detection claim made in Ref.~\cite{Isi:2019aib} was disputed by Ref.~\cite{Cotesta:2022pci} where a nearly identical time-domain analysis was reperformed (see also the reply Ref.~\cite{Isi:2022mhy}).
Applying the frequency-domain ringdown analysis originally presented in Ref.~\cite{Finch:2021qph}, we contribute to this discussion with a thorough reanalysis of the GW150914 data. 
This includes performing analyses with and without an overtone while considering different ringdown start times, as well as performing a noise injection study and studying the effects of different data sampling rates and frequency integration limits on our results.
Although the method used here differs significantly from previous time-domain analyses, we present our results in a way that makes it as easy as possible to compare with earlier work.
In conclusion, we do find tentative evidence for a ringdown overtone, but not at the high level of significance originally claimed in Ref.~\cite{Isi:2019aib}.

In order to be more quantitative, it is first necessary to be able to say clearly what it even means to ``detect a overtone''. 
Although intuitively obvious, it is not clear how to make this notion precise (this issue has previously been discussed in Ref.~\cite{Isi:2022mhy}). 
Several approaches have been suggested: looking to see if including the overtone improves the posterior on the remnant parameters (see Fig.~\ref{fig:mass_spin_post}); looking at the posterior on the overtone amplitude for a range of start times (see middle panel of Fig.~\ref{fig:overtone_amplitude}); computing the Bayes' factor in favor of an overtone (see bottom panel of Fig.~\ref{fig:overtone_amplitude}); and allowing the frequency of the second QNM to vary freely to see if the data prefers, or at least is consistent with, the expected Kerr value (see Figs.~\ref{fig:delta_f} and \ref{fig:other_QNMs}).
Although these are not all independent from one another, they all help shed light on which QNMs are present. 
The results of all of these tests can also be compared to results from a noise injection study.

As well as not being completely independent of each other, none of these tests are, by themselves, sufficient to justify a claim of a detection.
For example, one issue that has been raised is that the Bayes' factor can be made to take any value with a suitable adjustment to the prior range.
There are also conceptual problems regarding what it means to compare two models, neither of which is expected to fully describe the data. Here we are comparing the fundamental-only mode model (with a single QNM) to the overtone model (with two QNMs) when our firm prior belief is that the true signal should contain an infinite number of QNMs plus additional corrections (e.g.\ from nonlinearities in the merger, tails, and memory effects).

From the above discussion, it is clear that ringdown analyses are rather subtle. 
We think our frequency-domain method has some important advantages over what has been done before. 
For example, it marginalizes over the ringdown start time and sky position which is preferable to fixing these parameters (which potentially introduces systematic biases). 
Ideally, we should also marginalize over the uncertainties in the noise power spectral density (see, e.g.,\ Ref.~\cite{Cornish:2020dwh}) and detector calibration (see, e.g.,\ Ref.~\cite{2017PhRvD..96j2001C}) as part of a ringdown analysis. 
The ability to do this is, in principle, another benefit of the frequency-domain analysis approach used here as this can be done using techniques that are standard in the field.

We stress that while our results have been compared with those of previous time-domain studies, our frequency-domain method is rather different and therefore we do not expect to find perfect agreement. 
In contrast, the results of Refs.~\cite{Isi:2019aib, Cotesta:2022pci, Isi:2022mhy} are produced using essentially identical methods and should therefore be expected to agree exactly. 
The reason for the disagreement that is seen there is currently unknown and the subject of an ongoing investigation by both sets of authors.
It is vitally important for QNM science that all results are reproducible. To that end we have made all our data products and plotting scripts publicly available at Ref.~\cite{finch_eliot_zenodo}.

If QNMs are going to fulfill their promise for testing GR, fundamental physics and the Kerr metric hypothesis, then the community must be able to agree on standards for what it means to detect them and to be able to robustly quantify their significance. 
This field is still very young, and that there is already significant controversy regarding the QNM content of GW150914 and GW190521 is concerning, and we risk the situation becoming more confused with many more suitable events expected in O4.
And, as discussed in the introduction, this is a conceptual issue that will not be resolved with more observations, even at higher SNRs.
This issue needs input from the whole community; however, we suggest that (as a minimum) future claims of an overtone detection are accompanied by the investigations in Fig.~\ref{fig:mass_spin_post}, both panels of Fig.~\ref{fig:overtone_amplitude} and Fig.~\ref{fig:delta_f}.
That is, posteriors on the remnant properties with and without the overtone, posteriors on the overtone amplitude, a study of the Bayes' factor trends for different start times, and posteriors on deviations from Kerr when the overtone frequency is allowed to vary.

All data products and plot scripts associated with this work are made publicly available at Ref.~\cite{finch_eliot_zenodo}.


\begin{acknowledgments}
    We would like to thank the organizers and participants of the ringdown workshop held at the Center for Computational Astrophysics (Flatiron Institute, February 2022) where we had many useful discussions on this subject.
    We would like to thank Maximiliano Isi, Gregorio Carullo, Swetha Bhagwat, Ethan Payne and Juan Calderon Bustillo for comments on an early version of the manuscript.
    The computations described in this paper were performed using the University of Birmingham's BlueBEAR HPC service.
    Scientific color maps, available at Ref.~\cite{crameri_fabio_2021_5501399}, were used for the figures in this work.
    This research has made use of data or software obtained from the Gravitational Wave Open Science Center \cite{gwosc}, a service of LIGO Laboratory, the LIGO Scientific Collaboration, the Virgo Collaboration, and KAGRA. LIGO Laboratory and Advanced LIGO are funded by the United States National Science Foundation (NSF) as well as the Science and Technology Facilities Council (STFC) of the United Kingdom, the Max-Planck-Society (MPS), and the State of Niedersachsen/Germany for support of the construction of Advanced LIGO and construction and operation of the GEO600 detector. Additional support for Advanced LIGO was provided by the Australian Research Council. Virgo is funded, through the European Gravitational Observatory (EGO), by the French Centre National de Recherche Scientifique (CNRS), the Italian Istituto Nazionale di Fisica Nucleare (INFN) and the Dutch Nikhef, with contributions by institutions from Belgium, Germany, Greece, Hungary, Ireland, Japan, Monaco, Poland, Portugal, Spain. The construction and operation of KAGRA are funded by Ministry of Education, Culture, Sports, Science and Technology (MEXT), and Japan Society for the Promotion of Science (JSPS), National Research Foundation (NRF) and Ministry of Science and ICT (MSIT) in Korea, Academia Sinica (AS) and the Ministry of Science and Technology (MoST) in Taiwan.
    This document has been assigned LIGO document number P2200149.
\end{acknowledgments}


\bibliographystyle{apsrev4-1}
\bibliography{bibliography}


\appendix

\section{Sampling Frequency and Integration Limits}\label{app:fhigh}

Overtones ($n \geq 1$) have roughly the same real part of the frequency as the corresponding (i.e.\ with the same $\ell$ and $m$) fundamental model, but they have a shorter damping time. 
See, for example, Fig.~\ref{fig:other_QNMs}. 
This means that if a single, isolated, mode is viewed in Fourier space, the power spectrum is broader and contains significant power at higher frequencies.

It has been suggested \cite{WillMaxTGRtelecon} that overtone studies using highly downsampled data might suffer from aliasing effects due to the reduced Nyquist frequency.
Early ringdown studies, including those in Refs.~\cite{LIGOScientific:2016lio, Isi:2019aib, Isi:2022mhy}, generally used strain data that had been downsampled to $2\,$kHz. 
This was done for convenience and computational speed and was not originally anticipated to be a problem because the merger of GW150914 occurs at $\sim 200\,\mathrm{Hz}$, safely below the Nyquist frequency.

In this paper the $4\,$kHz data is used for all the analyses in the main text. 
Additionally, a frequency-domain log-likelihood with an upper integration limit of $f_{\rm high}=1000\,$Hz was used.
Our method is very different from the time-domain analyses, and do not expect our results to be sensitive to small changes in these choices.
To check that this is the case we have repeated the $N=1$ overtone analysis using the $16\,$kHz sampled data (obtained from \cite{gwosc}) and we find no significant changes in our results.
Using reweighting techniques (this time applied to the likelihood) we have also investigated the effect of changing the upper limit of integration in the likelihood. By re-evaluating the $N=1$ posterior chain on a likelihood with $f_\mathrm{ high}=1500\,$Hz and $2000\,$Hz, and then reweighting, we again find no significant changes in our results.

\section{Posteriors on the ringdown start time}\label{app:t0_posterior_prior}

\begin{figure}[t]
    \centering
    \includegraphics[width=\columnwidth]{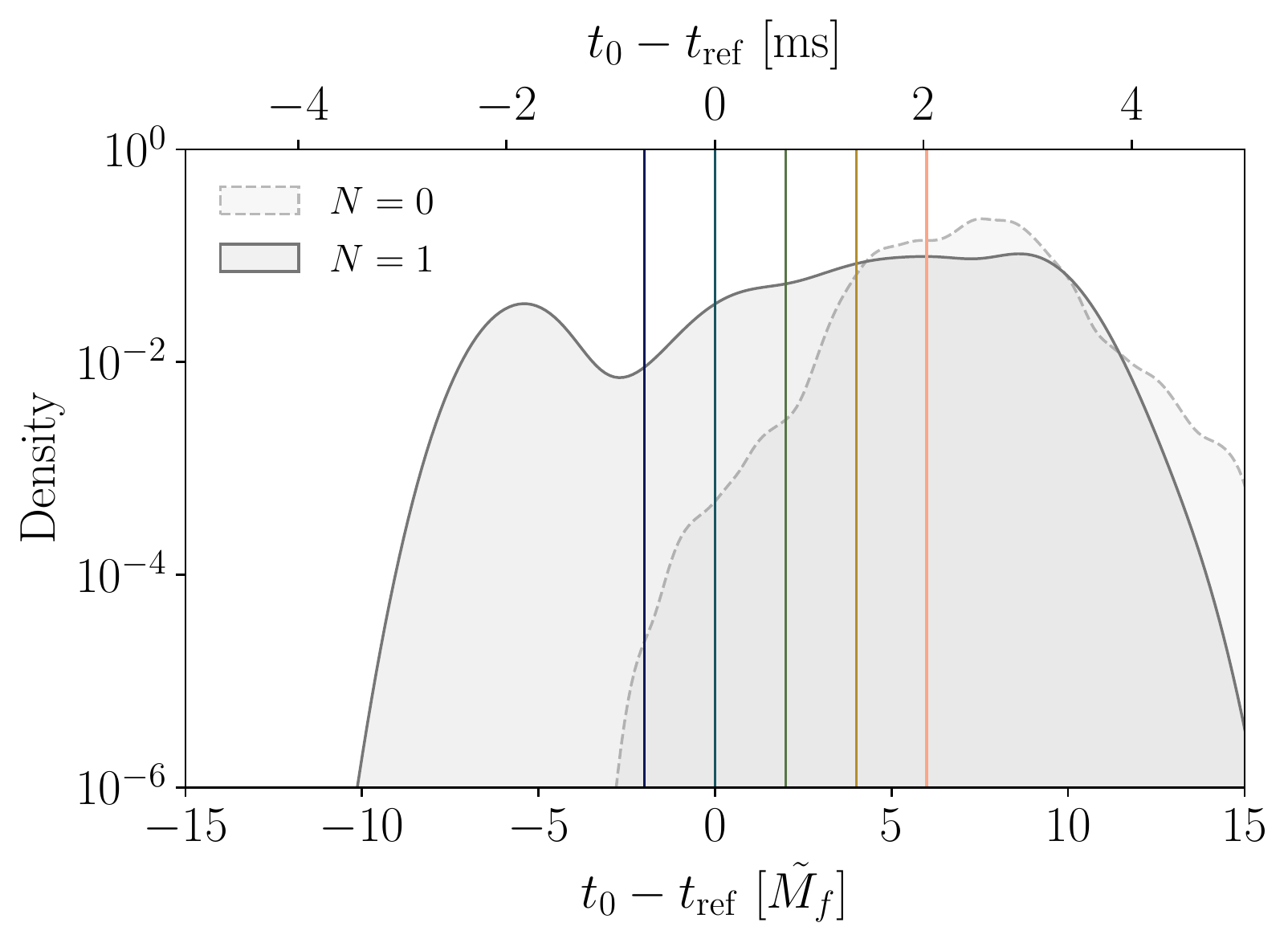}
    \caption{ \label{fig:t0_posterior}
        Posteriors on the ringdown start time obtained from our initial analysis using a flat prior over the range shown in the plot.
        Results are shown for the fundamental only ($N=0$) and overtone ($N=1$) analyses.
        Vertical colored lines show the locations of the means $\bar{t_0}$ of the narrow Gaussian priors used for the subsequent reweighting (see Fig.~\ref{fig:start_time}).
        The $N=1$ posterior has ample support across the entire range of interest, as required for the reweighting to remain accurate.
        The $N=0$ posterior has enough support everywhere except the $\bar{t_0}=-2\ \tilde{M_f}$ prior.
   }
\end{figure}

As described in Sec.~\ref{subsec:reweighting}, we initially perform Bayesian inference on the ringdown using a broad, flat prior on the ringdown start time parameter $t_0$. 
The posteriors on $t_0$ from both the $N=0$ and $N=1$ analyses are shown in Fig.~\ref{fig:t0_posterior}.
We do not consider these posteriors to be physically meaningful results because they were obtained with a prior that does not correctly describe our state of knowledge about when the ringdown should start.
These results are produced merely as an intermediate step in our analysis, before the reweighting was applied, and are shown here only to further illustrate the reweighting procedure described in Sec.~\ref{subsec:reweighting}.

In order for the subsequent reweighting step to be accurate, it is necessary for the posterior chains (particularly for the $N=1$ overtone analysis) to contain samples across the range of start times that we consider. 
For this reason, the \textsc{dynesty} sampler settings described in Sec.~\ref{sec:details} were chosen to ensure a large number of posterior samples were produced; we obtained 203697 and 218882 posterior samples from the $N=0$ and $N=1$ analyses respectively. 
This is sufficient for the reweighting to remain accurate everywhere except for the earliest start time in the $N=0$ analysis. 
This is the reason why this result is omitted from Fig.~\ref{fig:mass_spin_post}.

\section{Injection study}\label{app:inj}

Closely following the injection study performed in Ref.~\cite{Cotesta:2022pci}, we inject GW150914-like signals in the instrumental noise surrounding the true GW150914 event and reperform our overtone analysis ($N=1$).

The $\ell=2$ spin-weighted spherical harmonic of the numerical relativity simulation SXS:BBH:0305 \cite{2016CQGra..33x4002L, sxs_catalog} was used as the mock signal, scaled to a total mass of $72\,M_\odot$ and injected with a face-off orientation at a luminosity distance of $410\,\mathrm{Mpc}$. 
The sky position was taken to be $\alpha = 1.95\,\mathrm{rad}$, $\delta=-1.27\,\mathrm{rad}$.
This signal was injected into the data surrounding GW150914, such that the peak of the absolute value of the strain occurred at times $[-20, -15, -10, 5, 15, 20, 25, 30, 35, 40]\,\mathrm{s}$ relative to $t_\mathrm{ref}$.
These choices ensure the mock signal does not overlap with the real event.
Additionally, a zero-noise injection was performed for comparison. 

\begin{figure*}[t]
    \centering
    \includegraphics[width=2\columnwidth]{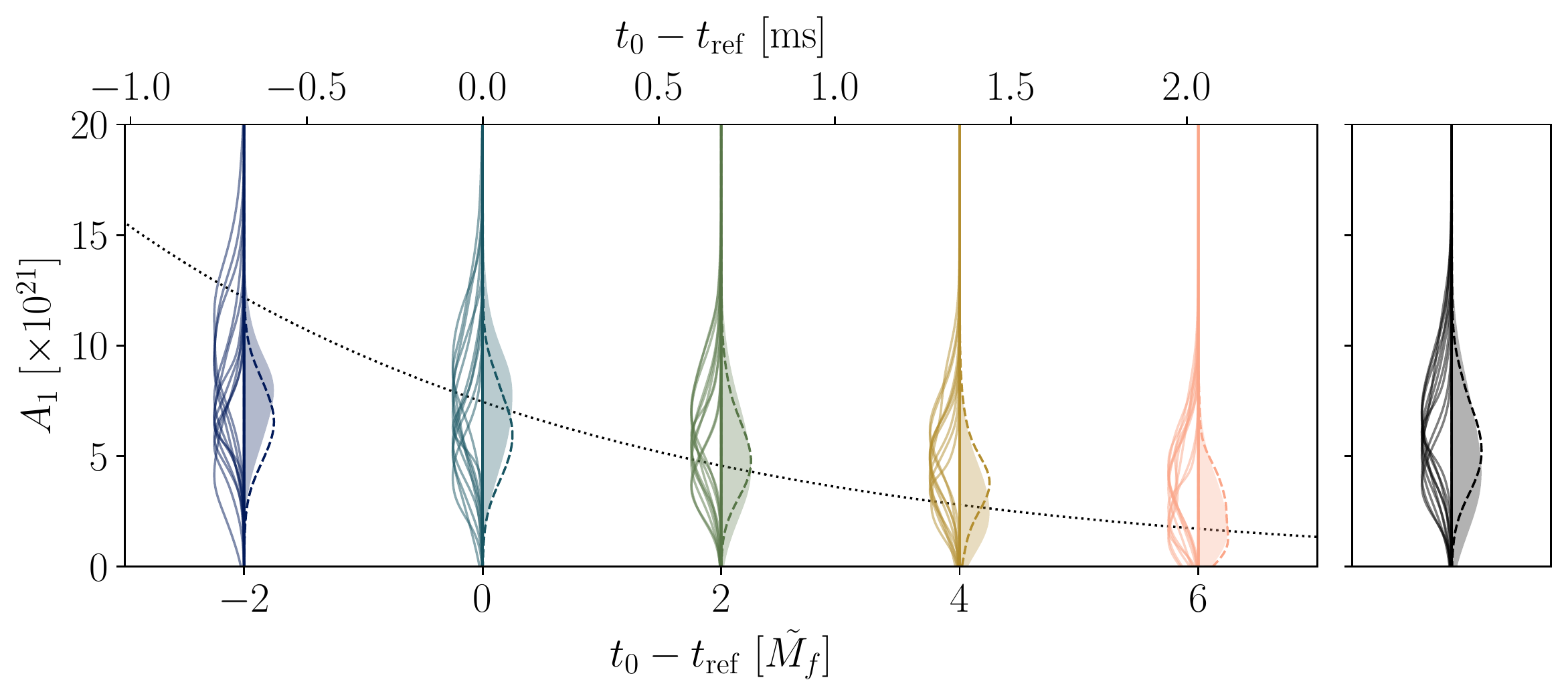}
    \caption{ \label{fig:injection_study}
        This is similar to the middle panel of Fig.~\ref{fig:overtone_amplitude} in the main text, but shows the posteriors on the overtone amplitude from the noise injection study.
        The different violin plots are for the different priors on the ringdown start time and the colors are the same as those used in Figs.~\ref{fig:start_time} and \ref{fig:overtone_amplitude}.
        On the right-hand side of each set of violin plots, the filled posterior shows the result obtained using the real GW150914 data (this is the same as what is plotted in Fig.~\ref{fig:overtone_amplitude}). 
        On the left-hand side are all the posteriors from the injection campaign, which indicate the spread in results due to different noise realizations.  
        Finally, the dashed lines on the right-hand side are the posteriors from the zero-noise injection.
        This plot is intended to be compared to Fig.~2 of Ref.~\cite{Cotesta:2022pci}, and Fig.~6 of Ref.~\cite{Isi:2022mhy}.
   }
\end{figure*}

We performed the frequency-domain ringdown analysis on these mock datasets using the same setup as was used for the real data and as described in Sec.~\ref{sec:details}.
This includes using the same PSD in the likelihood for all datasets.
We plot the resulting posteriors on the overtone amplitudes in Fig.~\ref{fig:injection_study}.
As with the real data, prior reweighting (see Sec.~\ref{subsec:reweighting}) has been used to show results for different choices of the ringdown start time prior.
We also investigated the Bayes' factors and found the same declining trend.

As expected, different noise realizations introduce some scatter into the results and we observe a spread in the locations of the maximum posterior values for the overtone amplitudes. 
However, this spread is consistent with the width of the posterior. 
The analysis described in the main text found only tentative evidence for the overtone, but there is no indication that this is overly effected by noise fluctuations.

\section{Wavelet posteriors}\label{app:W3}

The frequency-domain ringdown analysis method described in \cite{Finch:2021qph} and used here marginalizes over the early-time inspiral-merger signal using a flexible combination of sine-Gaussian wavelets (see Eq.~\ref{eq:wavelets}).
In the GW150914 analyses presented in this paper $W=3$ wavelets were used.
This choice was found empirically to be large enough to model the inspiral-merger signal without biasing the ringdown inference.
We have also verified that no strong correlations are observed between the wavelet and QNM parameters, and that further increasing the number of wavelets does not significantly affect the results for physically meaningful parameters (such as remnant parameters $M_f$ and $\chi_f$). 
These tests are described further in appendix A of Ref.~\cite{Finch:2021qph}.

\begin{figure*}[t]
    \centering
    \includegraphics[width=2\columnwidth]{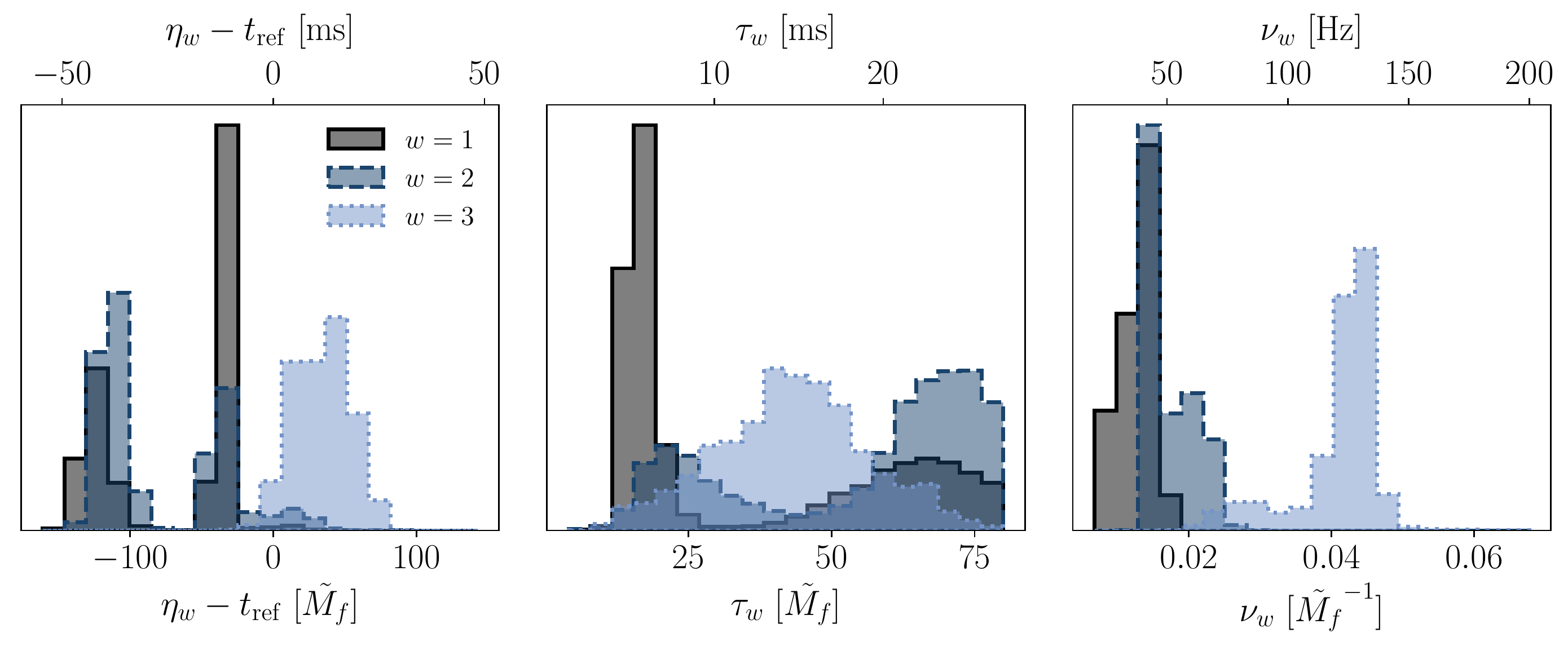}
    \caption{ \label{fig:wavelet}
        Posteriors on selected parameters for the $W=3$ wavelets used in the $N=1$ overtone analysis, reweighted using the IMRP $t_{\rm peak}$ prior on the ringdown start time.
        \emph{Left:} the wavelet central times, $\eta_w$.
        \emph{Middle:} the wavelet widths, $\tau_w$.
        \emph{Right:} the wavelet frequencies, $\nu_w$. 
        The index runs over values $w=1,\,2$ and $3$, where the numbering of the wavelets is chosen to enforce the ordering $\nu_{w}<\nu_{w+1}$.
        All plots use SI units on the upper $x$-axis and natural units on the lower $x$-axis. 
   }
\end{figure*}

The whitened strain posterior on the sum of these wavelets, together with the QNMs, can be seen in the early-time signal in Fig.~\ref{fig:waveform} where the fit to the data is seen to be excellent.
The wavelet parameters themselves are not physical; the wavelets are being used here solely to marginalize out the inspiral-merger. 
Nevertheless, in this appendix we show some additional posterior plots on the wavelet parameters, see Fig.~\ref{fig:wavelet}.
As expected, in order to describe the ``chirping'' inspiral signal, the wavelets naturally order themselves with their amplitudes and frequencies increasing with time. 

\end{document}